\documentstyle[titlepage]{article}
\newcommand{\R}{{\rm I\kern-2pt R}}

\def \and {\textnormal{ and }}

\def\R{\mathbb{R}}

\def\Tr{\textnormal{Tr}}
\def\bmbegin {\begin{bmatrix}}
\def\bmend {\end{bmatrix}}
\def\bme {\end{bmatrix}}
\def\bmb {\begin{bmatrix}}

\newtheorem{theorem}{Theorem}
\newtheorem{definition}[theorem]{Definition}

\newtheorem{proposition}[theorem]{Proposition}

\newtheorem{algorithm}[theorem]{Algorithm}

\newtheorem{example}[theorem]{Example}

\newtheorem{remark}[theorem]{Remark}

\begin{document}
\date{}
\title{Inversion of the Indefinite Double Covering Map}
\author{F. Adjei$^{\dagger}$, 
M. K. Dabkowski$^{*}$, S. Khan$^{*}$\& V. Ramakrishna$^{*}$\\
$\dagger$ : Department of Mathematics\\
Prairie View A\&M University\\
Prairie View, TX 77446, USA\\
{\it E-mail address}: {\texttt fradjei@pvamu.edu}\\[0.3cm]
$*$: Department of Mathematical Sciences\\
The University of Texas at Dallas\\
Richardson TX, 75080, USA\\
{\it E-mail addresses}: {\texttt mdab,Samreen.Sher,vish@utdallas.edu} 
}

\maketitle

\begin{abstract}
Algorithmic methods for the explicit inversion of 
the indefinite double covering maps are proposed. 
These are based on either the Givens decomposition or the 
polar decomposition of the given matrix in the proper, indefinite 
orthogonal group $SO^{+}(p, q)$.
As a by-product we establish that the 
preimage in the covering group, of a positive matrix in
$SO^{+}(p, q)$, can always be chosen to be itself positive 
definite. Inversion amounts to solving a poynomial system.
These methods solve this system by either inspection,
Groebner bases or by inverting the associated 
Lie algebra isomorphism and computing certain exponentials explicitly. 
The techniques are illustrated for 
$(p, q)\in \{(2,1), (2,2), (3,2), (4,1)\}$.\\[0.2cm]
\end{abstract}

\thispagestyle{plain}

\label{first}

\section{Introduction}\label{intro}
The double covers of the definite and indefinite orthogonal
groups, $SO^{+}(p, q)$ by the spin groups are venerable
objects. They arise in a variety of applications. The covering
of $SO(3)$ by $SU(2)$ is central to robotics. The group
$SO^{+}(2,1)$ arises in polarization optics, \cite{kim}. 
The group $SO^{+}(3,2)$
arises as the dynamical symmetry group of the 2D hydrogen atom and also
in the analysis of the DeSitter space time \cite{Gilmore,Kibler}. 
The group $SO^{+}(4,1)$
arises as the versor representation of the three dimensional conformal
group and thus has applications in robotics and computer vision,
\cite{perw}

In this work we provide explicit algorithms to invert these
double covering maps. 
More precisely,
given the double covering map,
$\Phi_{p,q}: {\mbox Spin}^{+}(p,q) \rightarrow SO^{+}(p, q)$ and
an $X\in SO^{+}(p, q)$ (henceforth called the {\it target}), 
we provide algorithms to compute the {\it matrices}
$\pm Y$, in the matrix algebra that the even subalgebra of 
${\mbox Cl} (p, q)$ is isomorphic to,
satisfying $\Phi_{p, q}(\pm Y) = X$.
One of our methods which works for all $(p,q)$, 
described in Remark (\ref{agnostic}),
finds the matrices in the preimage when ${\mbox Spin}^{+}
(p,q)$ is viewed as living in the matrix algebra that ${\mbox Cl}(p, q)$
is isomorphic to and this method trivially extends to the inversion of
the abstract double covering map. Our other methods, aid in finding
the preimage when ${\mbox Spin}^{+}
(p,q)$ is viewed as living in the matrix algebra that the even sublagebra
of ${\mbox Cl}(p, q)$ is isomorphic to. 
Since the even subalgebra typically consists
of matrices of lower size than ${\mbox Cl}(p, q)$ itself, such results
are of great interest. Naturally, such methods use the detailed
knowledge of the matrix forms of the various Clifford theoretic automorphisms
(such as grade, reversion), \cite{fpi}.
      
Our interest in finding $Y$ as a matrix, as opposed to an abstract
element of ${\mbox Cl}(p, q)$, stems from wishing to use specific matrix
theoretic properties of $Y$ (respectively $X$), 
together with an explicit matrix form of the direct map $\Phi_{p,q}$, to infer similar properties for $X$ (respectively $Y$). 
A classic example of this is the usage of the unit quaternions
(equivalently $SU(2)$) to find axis-angle representations of
$SO(3)$. In the same vein to compute the polar decomposition of
a matrix in $SO^{+}(3,2)$, it is easier to find that of its
preimage in the corresponding spin group, $Sp(4, \mathbf{R})$ and then
project both factors via $\Phi_{3,2}$. It is easy to show that the
projected factors constitute the polar decomposition of the original
matrix. Since the polar decomposition of a $4\times 4$ symplectic
matrix can even be found in closed form, \cite{jphysa}, this is
indeed a viable strategy. Similary, one can find the polar decomposition
of a matrix $X$ in $SO^{+}(4,1)$ in closed form, \cite{jciss} and this
can be then used to find the polar decomposition of $Y\in 
{\mbox Spin}^{+}(4,1)$. Since $Y$ is a $2\times 2$ matrix with
quaternionic entries and there is no method available for its
polar decomposition, \cite{leiba} (barring finding the polar decomposition
of an associated 
$4\times 4$ complex matrix), this is indeed useful.   
Similarly, block properties
of the preimage, $Y\in {\mbox Spin}^{+}(p, q)$,
viewed as a matrix, may provide useful information about $X$.
Such information is typically unavailable when finding $Y$ only as an  
abstract element of ${\mbox Cl} (p, q)$. Furthermore, one of 
the methods to be used
in this work, viz., the inversion of the linearization of the covering map, may
be used to also compute exponentials in the Lie algebras 
$so(p, q)$. 

There is literature on the problem being considered in this work.
For the case $(p, q) = (0, 3)$ [ or $(3, 0)$] this problem is
classical. The case $(0, 4)$ is considered in \cite{FourportI}.
The cases $(0, 5)$ and $(0, 6)$ are treated in \cite{JGSPII}.
The excellent work of
\cite{shiro} treats the general case with extensions to
the Pin groups, under a genericity assumption,
but finds the preimage in the abstract Clifford algebra via a formula
requiring the computation of several determinants. Section \ref{Section 1.1}
below provides a detailed discussion of the relation between
the present work and \cite{JGSPII,shiro}. In \cite{shiro2} an algorithm
is proposed, for the inversion of the abstract covering map,
but which requires the solution of a nonlinear equation
in several variables for which no techniques seem to be presented. 

\begin{remark}\label{CVA}
{\rm It is worth noting that even though the 
{\it abstract} map $\Phi_{p,q}$ is uniquely defined, as is 
the matrix algebra that ${\rm Cl}(p, q)$ is 
isomorphic to, the matrix form of $\Phi_{p, q}$ very much depends
on the matrices chosen to represent the one-vectors of ${\rm Cl}(p, q)$.
Similar statements apply to the explicit matrix forms of reversion and
Clifford conjugation. Indeed, even the group of matrices which
constitute ${\mbox Spin}^{+}(p, q)$ 
(within the matrix algebra that the 
even sublagebra of ${\rm Cl}(p,q)$ is isomorphic to) can change 
with the choice of one-vectors - though, of course, all 
these groups are isomorphic to each other. Thus, each abstract $\Phi_{p,q}$, 
in reality, 
determines many ``matrix", $\Phi_{p,q}$'s. 
Thus, our techniques are especially useful 
if one specific matrix form of $\Phi_{p,q}$ has been 
chosen in advance. 
This is illustrated via Example \ref{manyfold} later in this section.}
\end{remark}

With this understood, we will write $\Phi_{p,q}$ for the covering map,
with respect to the choice of one-vectors specified in advance.
Then $\Phi_{p, q}(Y)$ is a quadratic map in the entries of $Y
\in {\mbox Spin}^{+}(p,q)$. Given the target
$X\in {\rm SO^{+}}(p, q)$, we have to solve
this quadratic system of equations in several variables to obtain $Y$.
Solving this system is arduous in general. If $X$ can be factored as
a product of matrices $X_{j}$, then it is conceivable that these
systems become simpler for the individual $X_{j}$. Then using the
fact that $\Phi_{p, q}$ is a group homomorphism, one can synthesize
the solution for $X$ out of those for the $X_{j}$'s. This is the
technique employed in our earlier work, \cite{JGSPII}.
  
One standard choice of these simpler factors are standard Givens and hyperbolic Givens
rotations. These are matrices which are the identity except in a $2\times 2$
principal submatrix, where they are either a plane rotation or a hyperbolic
rotation. This decomposition of $X$ is constructive and thus leads
to a complete algorithmic procedure to find $Y$. For this it is important
that these Givens factors also belong to $SO^{+}(p, q)$. When $pq = 0$,
this is straightforward to show. The general case is less obvious
[since not all matrices of determinant $1$ in $SO(p, q)$ belong
to $SO^{+}(p,q)$] and is demonstrated in Section \ref{Section 2}
by displaying these Givens factors as exponentials of matrices in
the Lie algebra $so(p, q)$ (see Proposition \ref{inrightplace} in Section \ref{Section 2}).

As a second variant,
we also invoke the polar decomposition for some $(p, q)$ to
invert $\Phi_{p,q}$.  
This is aided by the remarkable fact that for certain $(p,q)$ (e.g., $(p,q)\in
\{(2,1), (3,1)\})$ $\Phi_{p,q}$ is inverted,
essentially by inspection, when the target in $SO^{+}(p, q)$ is positive
definite or special orthogonal. This circumstance may be used to
not only invert $\Phi_{p,q}$ but also compute the polar decomposition
of $X$ (or for that matter that of the preimage $\pm Y$)
without any eigencalculations. Once again, it is important to know that
both factors in the polar decomposition of an $X\in SO^{+}(p, q)$ belong
to $SO^{+}(p, q)$. It is known that for matrices in groups preserving
the bilinear form defined by $I_{p, q}$, both factors in the polar
decomposition do belong to the group also. However, it does not
immediately follow that the same holds for matrices in $SO^{+}(p, q)$.
Once again the issue at hand is that 
$SO(p, q)$ is not equal to $SO^{+}(p, q)$.
Inspite of this, as shown in \cite{jciss}, both factors in the polar
decomposition of a matrix in $
SO^{+}(p, q)$ also belong to $SO^{+}(p, q)$.

A related method to invert $\Phi_{p, q}$ is considered in this work.
Namely we invert instead the linearization $\Psi_{p,q}$ of
$\Phi_{p, q}$. The map $\Psi_{p, q}$ is easily inverted
since it is linear. For this method to be viable, however,
one has to first have an
explicit formula for the logarithm (within $so(p, q)$) of
$X$ or the $X_{j}$. 
Next, one has also to be able
to compute the exponential of this preimage. In Remark \ref{agnostic} we 
show that
both of these steps go through, 
when the target is decomposed via Givens decomposition \underline{regardless}
of the value of $(p, q)$.

In the interests of brevity, we illustrate only the cases
$(p, q) \in \{ (2,1)$, $(2,2)$, $(3,2)$, $(4,1)\}$. 
The $(2,1)$ case arises in polarization optics, for
instance. The $(2,2)$ case is
the simplest non-trivial case of the split orthogonal group. The $(3,2)$ case 
is of importance in the study of the hydrogen atom among other things.
The $(4,1)$ case is important in computer graphics.

Specifically, we will address
\begin{itemize}
\item The $(2,1)$ case by inspection and by
invoking the polar decomposition connection
aforementioned. The important $(3,1)$ case is also easily amenable
to this method, but surprisingly $(2,2)$ is not, \cite{jciss}.
\item The $(2,2)$ and $(3,2)$ cases via Groebner bases.
\item The $(4,1)$ case via linearizing $\Phi_{4,1}$. We will provide
results both when $X$ is decomposed using Givens factors and when it
is decomposed using polar decomposition.
\end{itemize}

\subsection{Relation to Work in the Literature}\label{Section 1.1}
In this section the relation of the present work with
\cite{JGSPII,shiro} is dicusssed.

The relation between \cite{JGSPII} and the present work is as follows.
In \cite{JGSPII} the concern is with inversion for the $(0, 5)$ and
$(0, 6)$ cases. Since the polar decomposition of an orthogonal matrix
is trivial, it does not help at all with the task of inversion.
On the other hand, in this work it plays a significant role precisely
because the polar decomposition for the $(p, q), pq\neq 0$ case is no
longer trivial. Therefore, when the inversion of a positive definite
target in $SO^{+}(p, q)$ can be carried out efficiently, it becomes
even more useful than when the target is an ordinary or hyperbolic
Givens matrix, since the number of Givens factor grows rapidly with $n = p+q$.

Next the relation to \cite{shiro} is dicussed. 
Signifcant portions of the present work were completed in late 2016, 
\cite{ThesisAdjei}. As this paper was being written up, we became
aware of the 2019 paper \cite{shiro}. 
In \cite{shiro}, an elegant solution is provided
for inverting the {\it abstract} (as opposed to the matrix)
map $\Phi_{p, q}$, under a generic
condition on $X$, not required by our work. The solution in
\cite{shiro} is a generalization of a method proposed in \cite{hestenes}
for the $(3,1)$ case. 
This formula is as follows. Let $X\in SO^{+}(p, q)$
and define the element $M$ of ${\mbox Cl}(p, q)$ via
\[
M = \sum_{\alpha , \beta} {\rm det}(X_{\alpha , \beta})e_{\alpha}
(e_{\beta})^{-1}.
\]
Here $\alpha = \{i_{1}, \ldots, i_{k}\}$ and $\beta =
\{j_{1}, \ldots, j_{k}\}$ are subsets of $\{1,2, \ldots , n\}$ of equal
cardinality (including the empty set), $X_{\alpha , \beta}$ is the square 
submatrix
of $X$ located in rows indexed by $\alpha$ and
and columns indexed by $\beta$,  
$e_{\alpha} = e_{i_{1}}e_{i_{2}}\ldots e_{i_{k}}$ and similarly
for $e_{\beta}$. Here $e_{l}$ is the $l$-th basis one-vector in
the abstract Clifford algebra $ {\mbox Cl}(p, q)$. $(e_{\beta})^{-1}$ is the
inverse of $e_{\beta}$ in ${\mbox Cl}(p, q)$.

It is assumed that $MM^{rev}\neq 0$ in ${\mbox Cl}(p, q)$. 
This is the aforementioned
genericity assumption. Then $\Phi_{p, q}(\pm Y)
= X$,
where
\[
Y = \frac{M}{\sqrt{MM^{rev}}}
\]
Here $M^{rev}$ is the reversion of $M$.

\begin{remark}\label{nuis}
{\rm The following nuances, besides the genericity condition
$MM^{\rm rev}\neq 0$, of this
formula need attention
\begin{itemize} 
\item[i)]  The principal burden in implementing the formula in \cite{shiro} is that one has to 
compute all the minors of $X$. If we ally this formula with 
one innovation of the current work, namely decomposing $X$ into
hyerbolic and usual Givens rotations, then a significant reduction in
complexity in implementing the formula in \cite{shiro} can be expected.
Indeed, the number of non-zero minors of a hyperbolic or standard Givens
is much lower than that for a general $X$. However, from many viewpoints,
it is still emphatically not true that if $X$ is a Givens rotation then only
principal minors are non-zero, and thus still several determinants
have to be computed. For instance, consider
\[
X = \left (\begin{array}{ccc}
a & 0 & b\\
0 & 1 & 0\\
b & 0 & a
\end{array}
\right ), a^{2} - b^{2} = 1
\]
Thus $X$ is a hyperbolic Givens rotation in $SO^{+}(2,1)$.
Then, for instance, the following non-principal $2\times 2$ minors are non-zero:
$\{ (1,2)$, $(2,3)\}$, $\{(2,3)$, $(1,2)\}$.
Hence, even when the target $X$ is sparse, such as a Givens matrix, one
has to calculate several determinants. 

\item[ii)] Furthermore, due to the involvement of several determinants, the formula
obtained for $Y$ often is quite elaborate and occludes the ``half-angle''
nature of the inversion even when $X$ is simple - 
see Example \ref{manyfold} below for an illustration
of this issue.
\item[iii)] The formula only finds $Y$ as an element of the abstract
Clifford algebra ${\mbox Cl}(p, q)$. Our methods also provide 
such an inversion, without
the need for determinants, but by 
using Givens decompositions - see Remark \ref{agnostic}. 
Of course, by using specific matrices
constituting a basis
of one-vectors for ${\mbox Cl}(p, q)$, $Y$ can be recovered as a matrix.
The matrix $Y$ thereby obtained, even though an even vector, will live in
${\mbox Cl}(p, q)$ which is typically an algebra of matrices of {\it larger}
size than the matrix algebra constituting the even subalgebra.
This is due to the very nature of the formula.
Thus, for instance this formula will yield, for the case $(p,q)
= (3,2)$, $Y$ as a $8\times 8$ matrix, even though the covering
group consists of symplectic matrices of size $4$. To get around this
one has to know how to embed the even subalgebra in 
${\mbox Cl}(p, q)$, \cite{fpi}.
In effect, one has to compute the matrix form of the grade involution.
This limitation is thus due to not having at one's
disposal an explicit matrix form for $\Phi_{p, q}$, when using
the formula in \cite{shiro}.
     
\item[iv)] Next the matrix form of 
$\pm Y$ very much depends on the basis 
of one-vectors. 
Without this caveat, one can find different find different matrices, 
$Y$, with $\pm Y$ projecting to the same $X$. 
This matter is illustrated in Example \ref{manyfold}.

\item[v)] Other steps in this method such as finding the reversion of $M$ 
can, in principle, be performed without having to resort to
finding reversion as an explicit automorphism of the matrix algebra that
${\mbox Cl}(p, q)$ is isomorphic to. However, as $p + q$ grows,
it is more convenient to work with a concrete matrix form of
reversion, such as those in \cite{fpi}). Indeed $MM^{rev}$ is proportional
to the identity and thus, if a matrix form of $M$ (and $M^{rev}$)
is available, then one needs to only compute the trace of $MM^{rev}$.
These issues are all mitigated when the methods being proposed here are used,
since our methods make systematic use of the structure of the 
matrix form of the map $\Phi_{p,q}$, whereas
this is not the case in \cite{shiro}.
\end{itemize}
}
\end{remark}            

\begin{example}\label{manyfold}
{\rm Consider $\Phi_{1,1}$. Its inversion, is of course, trivial.
However, it illustrates Remark \ref{CVA} and also the caveats ii) and iv)
above about the usage of the formula in \cite{shiro}.

Let us use the basis ${\mathbf B}_{1} = \{\sigma_{x}, i\sigma_{y}\}$ for the one-vectors of ${\rm Cl}(1,1)
\simeq M(2, \mathbf{R})$. Incidentally,
this is the canonical basis that the constructive proof of the 
isomorphism between ${\rm Cl}(p+1, q+1)$ and $M(2, {\rm Cl}(p, q))$ 
naturally yields. Then
${\mbox Spin^{+}(1,1)}$ is realized as
\begin{equation}\label{First}
\begin{array}{rcl}
{\mbox Spin^{+}}(1,1) &=& \{ \left (\begin{array}{cc}
\alpha  & 0\\
0 & 1/\alpha
\end{array}
\right ) ; 
\alpha\neq 0\}.
\end{array}
\end{equation}

Now
\[\Phi_{1,1}
[\left (\begin{array}{cc}
\alpha  & 0\\
0 & 1/\alpha
\end{array}
\right )]= \frac{1}{2}\left (\begin{array}{cc}
\frac{\alpha^{2} + 1}{\alpha^{4}} & \frac{\alpha^{2} - 1}{\alpha^{4}}\\
\frac{\alpha^{2} - 1}{\alpha^{4}} & \frac{\alpha^{2} + 1}{\alpha^{4}}
\end{array}
\right )\cdot\]
Let 
$X = \left (\begin{array}{cc}
a & b\\
b & a
\end{array}
\right )$ be a target matrix is $SO^{+}(1,1)$. Here
\[
 a = \cosh (x),\quad b = \sinh (x)
\] 
Directly solving for $\alpha$ from the quadratic system obtained 
from
\[X=\Phi_{1,1} [ \left (
\begin{array}{cc}
\alpha & 0\\
0 & 1/\alpha
\end{array}
\right )]\]
one recovers 
\begin{equation}
\label{CVAfirst}
Y = \pm
\left (\begin{array}{cc}
e^{x/2} & 0\\
0 & e^{-x/2} 
\end{array}
\right ).
\end{equation}
The ``half-angle'' aspect of the covering map is manifest in this formula.
 
Only after some algebra, is this also
the solution 
yielded by the formula of \cite{shiro}. Specifically
\[
M = (2 + 2a)1 + 2be_{2}e_{1} =  \left (\begin{array}{cc}
2 + 2a + 2b & 0\\
0 & 2 + 2a-2b 
\end{array}
\right )
\]
since $e_{2}e_{1}
= \sigma_{z}$ if we use $\mathbf{B}_{1}$ as the basis of one-vectors for
${\rm Cl}(1,1)$. Next, a calculation shows $MM^{rev} = (8 + 8a)1$. So, it follows that
\begin{equation}\label{CVAsecond} 
Y = \left (\begin{array}{cc}
\frac{2 + 2a + 2b}{\sqrt{8+8a}} & 0\\
0 & \frac{2 + 2a - 2b}{\sqrt{8+8a}}
\end{array}
\right )
\end{equation}
which, after further manipulations, coincides with 
Equation (\ref{CVAfirst}).
Thus, even in this simple case, it is seen that if one in interested
in a symbolic expression for $Y$ as a function of the
entries of $X$, then Equation (\ref{CVAsecond}) 
is more complicated that Equation (\ref{CVAfirst}), even though
they are equivalent. Next, we could also have used
${\mathbf B}_{2} = \{ \sigma_{z}, i\sigma_{y}\}$, as the basis of
one-vectors. Naively applying the formula
in \cite{shiro} would then naturally lead to
$Y$ being a linear combination of
$I_{2}$ and $\sigma_{x}$, which is inconsistent with Equation (\ref{First}).
The resolution is that with $\mathbf{B}_{2}$ as the choice of
the basis of one-vectors, ${\mbox Spin}^{+}(1,1)$ is just
\begin{equation}
\begin{array}{rcl}
{\mbox Spin}^{+}(1,1) &=& \{ \left (\begin{array}{cc}
	a & b\\
	b & a
\end{array}
\right ) ; 
a^{2} - b^{2} = 1\}.
\end{array}
\end{equation}
}
\end{example}

\subsection{Organization of the Paper}\label{Section 1.2}
The balance of the paper is organized as follows. In Section \ref{Section 2.1}, we
record notation used throughout the work, Section \ref{Section 2.2} records definitions
from Clifford algebras, especially that of the covering map and its
linearization. Section \ref{Section 2.3} discusses matrices with quaternionic entries
and their complex representations. We draw attention to Remark
\ref{Hproperties} and Remark \ref{sundry}. Section \ref{Section 2.4} shows that
inverse images of positive definite matrices can be chosen to
be positive definite matrices in the covering group - 
see Proposition \ref{InvisPosDefinite}. This result
assumes that the basis of one-vectors chosen satisfy two properties
(called {\bf BP1} and {\bf BP2}). In the Appendix we show that every
real Clifford algebra has at least one such basis. Section \ref{Section 2.4} discusses the polar decomposition and the Givens 
decompositions. We draw attention to
\begin{itemize}
\item Remark \ref{PDforn1} which discusses
a constructive algorithm, Algorithm \ref{PDbyRotation} in Section \ref{Section 3}, for the polar decomposition in certain indefinite orthogonal groups.
\item Remark \ref{FindingLog}.
\item Proposition \ref{inrightplace}
which shows that Givens matrices indeed belong to 
$SO^{+}(p,q)$, and 
\item Example \ref{Givens} which illustrates how
any matrix in the indefinite orthogonal groups can be factored constructively into Givens factors.
\end{itemize}
Section \ref{Section 3} is devoted to the inversion of $\Phi_{2,1}$. In this section
we use the polar decomposition and the key observation is that if
$P\in {\rm SO^{+}}(2,1)$ is positive definite, then its preimage(s) under
$\Phi_{2,1}$ can be found by inspection. In sections \ref{Section 4} and \ref{Section 5} we directly
invert $\Phi_{2,2}$ and $\Phi_{3,2}$ by solving the systems
of quadratic equations
when the target is a Givens factor. In Section \ref{Section 6}, we invert $\Phi_{4,1}$
by first inverting the linearization $\Psi_{4,1}$ and then showing
that the exponential of the matrix in the Lie algebra of the spin group,
thereby obtained, admits a closed-form expression. We demonstrate this
when the target is decomposed using Givens factors or when the polar
decomposition is employed for the same purpose. In particular this
provides an algorithm for finding the polar decomposition of certain
$2\times 2$ quaternionic matrices without having to find that of
the associated $4\times 4$ complex matrix. 
We draw attention to Remark \ref{agnostic}
of this section which shows that 
linearization provides a viable method to invert 
the {\it abstract} covering map, when used 
in conjunction with Givens decompositions.
The last section offers conclusions.

\section{Preliminaries}\label{Section 2}
\subsection{Notation}\label{Section 2.1}
We use the following notation throughout

\begin{description}
\item[N1] $\mathbf{H}$ is the set of quaternions. Let $K$ be an associative algebra. Then
$M(n, K)$ is just the set of $n\times n$ matrices with entries in
$K$. For $K=\mathbf{C},\mathbf{H}$ we define $X^{\ast }$ as the matrix obtained by
performing entrywise complex (respectively quaternionic) conjugation first, and
then transposition. For $K=\mathbf{C}$, $\bar{X}$ is the matrix obtained by performing entrywise complex
conjugation.

\item[N2] The Pauli Matrices are
\[
\sigma _{x}=\sigma _{1}=\left( 
\begin{array}{cc}
0 & 1 \\ 
1 & 0%
\end{array}%
\right); \sigma _{y}=\sigma _{2}=\left( 
\begin{array}{cc}
0 & -\i \\ 
\i & 0
\end{array}
\right) ; \sigma _{z}=\sigma _{3}=\left( 
\begin{array}{cc}
1 & 0 \\ 
0 & -1%
\end{array}
\right). 
\]
\end{description}

\subsection{Preliminary Observations}\label{Section 2.2}
We will begin with informal definitions of the notions of one and
two-vectors for a Clifford algebra, which is sufficient for the purpose of
this work. The texts \cite{pertii,portei} are excellent sources for more
precise definitions in the theory of Clifford algebras. The same texts
also contain precise definitions of the various automorphisms
(such as grade, Clifford conjugation and reversion).

\begin{definition}
\label{One-vector}{\rm Let $p,$ $q$ be non-negative integers with $p+q=n$. A
collection of matrices $\left\{ X_{1},\ldots ,X_{p},
X_{p+1},\ldots ,X_{p+q}\right\},$ 
with entries in $\mathbf{R}, \mathbf{C}$ or $\mathbf{H}$ 
is a basis of one-vectors for the Clifford algebra ${\mbox Cl}( p,q) $ if

\begin{description}
\item[1.] $X_{i}^{2}= I$, for $i=1, 2,..., p,$ where $I$ is the identity matrix of the appropriate size $($this size is typically different from $n)$
\item[2.] $X_{i}^{2} =-I$, for $i=p+1, p+2, ..., p+q$
\item[3.] $X_{i}X_{j}=-X_{j}X_{i}$, for $i\neq j$; $i,j=1, 2,..., n$.
\end{description}

A one-vector is just a real linear combination of the $X_{i}$'s, $i=1, 2, ...,n$. Similarly, a two-vector is a real linear combination of the
matrices $X_{i}X_{j}$, $i<j$, $i,j=1,2,...,n$. Analogously, we
can define three, four, ..., $n$-vectors, etc. 
${\mbox Cl}(p, q)$ is just a real linear combination of $I$, one-vectors,
..., $n$-vectors.}
\end{definition}

\begin{definition}\label{basics}
${\mbox Spin^{+}}( p, q)$ {\rm is the connected component of 
the identity of the  collection of
elements $x$ in ${\mbox Cl}(p, q)$ satisfying the
following requirements: i)$x^{gr}=x$, i.e., $x$ is even
(here $x^{gr}$ is the grade involution applied to $x$), $ii)$
$xx^{cc}=1$ (here $x^{cc}$ is the Clifford conjugate
of $x$) and $iii)$ For all one-vectors $v$ in ${\mbox Cl}( 0, n)$, 
$xvx^{cc}$ is also a one-vector. The last condition, in the
presence of the first two conditions, is 
known to be superfluous for $p+q\leq 5$, \cite{pertii,portei}.
}
\end{definition}

\begin{definition} {\rm Let $n= p + q$. Denote by $I_{p,q} = I_{p}\oplus (-I_{q})$.
Then ${\rm SO}(p, q) = \{X\in M(n , {\mathbf R}): 
X^{T}I_{p,q}X = I_{p,q}; \det (X) = 1\}$.
$SO^{+}(p, q)$ is the connected component of the identity in $SO(p, q)$. 
Unless
$pq= 0$, $SO^{+}(p, q)$ is a proper subset of $SO(p, q)$. 
The Lie algebra of $SO^{+}(p, q)$ is denoted by $so(p,q)$ and
it is described by 
\[
so(p, q) = \{X: X\in M(n, {\mathbf R}), X^{T}I_{p, q}= -I_{p,q}X\}
\]
} 
\end{definition}

\begin{definition}\label{PhiPsiDefine}
{\rm The map $\Phi_{p,q}: {\mbox Spin^{+}}( p, q)\rightarrow SO^{+}(p, q)$ 
sends $x\in {\mbox Spin^{+}}( p, q)$ to the matrix of
the linear map $v\rightarrow xvx^{cc}$, where $v$ is a one-vector
with respect to a basis $\{X_{1}, \ldots , X_{p}, X_{p+1}, \ldots , 
X_{p+q}\}$ of the space of one-vectors. ${\mbox spin}^{+}(p, q)$ is its
Lie algebra and is known to equal the space of bivectors of ${\mbox Cl}(p, q)$.
It is further known that
$\Phi_{p, q}$ is a group homomorphism with kernel $\{\pm I\}$.
We denote by $\Psi_{p, q}$ the linearization of $\Phi_{p, q}$.
Thus, $\Psi_{p, q}$ sends an element $y\in {\mbox spin}^{+}( p, q)$ 
to the matrix of the linear map $v\rightarrow yv - vy$. $\Psi_{p, q}$
is a Lie algebra isomorphism from ${\mbox spin}^{+}( p, q)$ to 
$so(p, q)$.}             

\end{definition}

\subsection{Quaternionic and $\theta _{\mathbf{H}}$ Matrices}\label{Section 2.3}

\noindent Next, to a matrix with quaternion entries will be associated a
complex matrix. First, if $q\in {\mathbf H}$ is a quaternion, it can be
written uniquely in the form $q=z+ w j$, for some $z, w\in { \mathbf C}$. 
Note that $j\eta =\bar{\eta}j$, for any $\eta \in \mathbf{C}$. 
With this at hand, the following construction associating 
complex matrices to matrices with quaternionic entries is useful.

\begin{definition}
{\rm Let $X\in M(n,\mathbf{H})$. By writing each entry $x_{pq}$ of $X$
as
\[
x_{pq}=z_{pq}+w_{pq}\bf{j},\ z_{pq},w_{pq}\in \mathbf{C}
\]
we can write $X$ uniquely as $X=Z+W{\bf j}$ with 
$Z,W\in M( n,\mathbf{C})$. Associate to $X$ the 
following matrix $\theta _{\mathbf{H}}(X)\in 
M(2n,\mathbf{C})$
\[
\theta _{\mathbf{H}}(X)=\left( 
\begin{array}{cc}
Z & W \\ 
-\bar{W} & \bar{Z}%
\end{array}%
\right). 
\]
}
\end{definition}

\begin{remark}
\label{IntertwineConjugation} {\rm Viewing an $X\in M( n,\mathbf{C})$
as an element of $M(n,\mathbf{H})$ it is immediate
that $j X=\bar{X}j$, where $\bar{X}$ is entrywise complex
conjugation of $X$.
}
\end{remark}

\begin{definition}
{\rm A $2n\times 2n$ complex matrix of the form 
$\left ( 
\begin{array}{cc}
Z & W \\ 
-\bar{W} & \bar{Z}%
\end{array}%
\right )$ is said to be a $\theta _{\mathbf{H}}$ matrix.
}
\end{definition}
 
\noindent Next some useful properties of the map 
$\theta _{\mathbf{H}}:M(n, {\mathbf H})\rightarrow M(2n, {\mathbf C})$
are collected.

\begin{remark}\label{Hproperties} 
Properties of $\theta _{\mathbf{H}}$

{\rm \begin{description}
\item[i)] $\theta _{\mathbf{H}}$ is an $\mathbf{R}$-linear map

\item[ii)] $\theta _{\mathbf{H}}(XY)=
\theta _{\mathbf{H}}(X)\theta _{\mathbf{H}}(Y)$

\item[iii)] $\theta _{\mathbf{H}}(X^{\ast })=[\theta _{\mathbf{H}}(X)]^{\ast
}$. Here the $\ast $ on the left is quaternionic Hermitian conjugation,
while that on the right is complex Hermitian conjugation.
                                                             
\item[iv)] $\theta _{\mathbf{H}}(I_{n})=I_{2n}.$

\end{description}
}
\end{remark}

\begin{remark}\label{sundry}
{\rm In this remark we will collect some more facts concerning
quaternionic matrices. 
\begin{enumerate}
\item If $X, Y\in M(n,{\mathbf H})$ then
it is not true that ${\mbox Tr}(XY) = {\mbox Tr}(YX)$.
However, ${\mbox Re} \ ({\mbox Tr}(XY)) = {\mbox Re}({\mbox Tr}(YX))$.
Therefore, the following version of cyclic invariance of trace holds
for quaternionic matrices
\[
{\mbox Re}[{\mbox Tr}(XYZ)] =
{\mbox Re}[{\mbox Tr}(YZX)] = {\mbox Re}([{\mbox Tr}(ZXY)]
\]
\item Let $X$ and $Y$ be square quaternionic matrices.
Then ${\mbox Tr}(X\otimes Y) = {\mbox Tr}(X){\mbox Tr}(Y).$
Furthermore, if at least one of $X$ and $Y$ is real, then
\[
{\mbox Re}({\mbox Tr}(X\otimes Y))
= {\mbox Re}({\mbox Tr}(X)){\mbox Re}({\mbox Tr}(Y))
\]

\item $X = Z + Wj$ is Hermitian iff $Z$ is Hermitian (as a complex
matrix) and $W$ is skew-symmetric. This is, of course, equivalent
to $ \theta _{\mathbf{H}}(X)$ being Hermitian as a complex matrix.

\item If $X$ is a square quaternionic matrix, we define
$${\mbox Exp}(X) = I_{n} + X + \frac{X^{2}}{2!} + \ldots .$$
Then $\theta _{\mathbf{H}}({\mbox Exp}(X)) = {\mbox Exp} 
( \theta _{\mathbf{H}}(X))$.

\item If $X$ is a square quaternionic matrix, it is positive definite
if $q^{*}Xq > 0$, for all $q \in \mathbf{H}$. This is equivalent
to $ \theta _{\mathbf{H}}(X)$ being a positive definite complex matrix.

\item Putting the last two items together we see that if $X\in
M(n,\mathbf{H})$ is Hermitian, then ${\mbox Exp}(X)$ is positive definite.

\end{enumerate}
}
\end{remark}
  
We next prove a very useful result, Proposition \ref{InvisPosDefinite},
 which ensures that one preimage
in ${\mbox Spin^{+}}(p, q)$ of a positive definite matrix in
$SO^{+}(p, q)$ is also positive definite. In light especially of \ \\
Remark \ref{manyfold}, it must be stressed that it is being assumed
in Proposition \ref{InvisPosDefinite} that the basis, $\mathbf{B}$, of
one-vectors for ${\mbox Cl}(p, q)$ being used satisifies the following two
properties
\begin{itemize}
\item {\bf BP1} If $V\in \mathbf{B}$, the basis of one-vectors 
being used, then
\begin{equation}
\label{BP1}
V_{i}^{*} =\pm V_{i}
\end{equation}

\item {\bf BP2} The matrices in $\mathbf{B}$ are orthogonal with respect to
the trace inner product. Specifically, if $\mathbf{B}$ consists of real or
complex matrices then
\begin{equation}
\label{BP2}
{\mbox Tr}(U^{*}V) = 0, \forall U, V \in \mathbf{B}, U\neq V
\end{equation}
and if $\mathbf{B}$ contains quaternionic matrices then
\begin{equation}\label{BP3}
{\mbox Re}({\mbox Tr}(U^{*}V)) = 0, \forall U, V \in \mathbf{B}, U\neq V.   
\end{equation}   
\end{itemize}    
\begin{proposition}\label{InvisPosDefinite}
{\rm Let $P\in SO^{+}(p,q)$ be positive definite. Let
$\mathbf{B}$ be a set of matrices serving as a basis of one-vectors
for ${\mbox Cl}(p, q)$ which satisfy both {\bf BP1} and {\bf BP2}.
Then there is a unique positive definite $Y\in {\mbox Spin}^{+}(p, q)$
with $\Phi_{p, q} (Y) = P$.}
\end{proposition}

\noindent {\bf Proof:}
As shown in \cite{jciss}, there is a symmetric $Q\in so(p,q)$
such that
${\mbox Exp}(Q) = P$. Let $\Psi_{p, q}$ be the linearization of
$\Phi_{p, q}$. We will show that the (unique) preimage $A$
of $Q$ with respect to $\Psi_{p, q}$ is
Hermitian. Therefore, from the formula $\Phi_{p, q}[{\mbox Exp}(A)]
= {\mbox Exp} [\Psi_{p, q} (A)]$, it follows that
if we denote by $Y = {\mbox Exp}(A)$, then $\Phi_{p, q} (Y) = P$.
Since $A$ is Hermitian, it follows that 
$Y = {\mbox Exp}(A)$ is positive definite (where, in the event $A$ is quaternionic we invoke the last item of Remark \ref{sundry}).

Let us now show that $A = \Psi^{-1}_{p,q}(Q)$ is
Hermitian. First, suppose that $\mathbf{B}$ consists
of real or complex matrices. Since $\mathbf{B}$ satisfies 
{\bf BP1} and {\bf BP2} we have 
\begin{itemize}
\item 1) If $A \in {\mbox spin}^{+}(p, q)$, then
$A^{*}\in {\mbox spin}^{+}(p, q)$ also. Indeed the typical
element of ${\mbox spin}^{+}(p, q)$ is a real linear combination
of the bivectors $V_{k}V_{l}, k < l$. Since
$(V_{k}V_{l})^{*} = \pm V_{k}V_{l}$
(using the fact that the $V_{i}$s anticommute), it follows
that $A^{*}$ is also a real linear combination of
the bivectors and is thus in ${\mbox spin}^{+}(p, q)$ also.

\item 2) $\Psi_{p,q}(A^{*}) =
[\Psi_{p, q}(A)]^{*}$. To see this note that 
the $(i, j)$th entry of the matrix $\Psi_{p, q}(A)$ equals, due
the $V_{i}$'s being orthogonal with respect to the trace inner product
\[
[\Psi_{p,q}(A)]_{ij} = {\mbox Tr}[V_{i}^{*} (AV_{j} - V_{j}A)]
= {\Tr}[A (V_{j}V_{i}^{*} - V_{i}^{*}V_{j})]
\]
(where we used the cyclic invariance of trace).

Similarly the $(j, i)$ entry of $\Psi_{p,q}(A^{*})$ equals
${\mbox Tr} [A^{*} (V_{i}V_{j}^{*} - V_{j}^{*}V_{i})]$.
But this equals the complex conjugate of 
${\mbox Tr} [(V_{j}V_{i}^{*} - V_{i}^{*}V_{j})A]$, which again
by cyclic invariance of trace, equals $\overline{\Psi_{p,q}(A)_{ij}}$.
\end{itemize}

If $\mathbf{B}$ contains quaternionic matrices then the above argument
goes through verbatim if we replace ${\mbox Tr}$ by ${\mbox Re}\ 
{\mbox Tr}$
in light of item 1) of Remark \ref{sundry}.

So (as all $\Psi_{p,q}(A)$ are real), if $\Psi_{p, q}(A)$
is symmetric then,  in light of $\Psi_{p, q}$ being a vector
space isomorphism, it follows that $A = A^{*}$ and hence $A$ is Hermitian and
$Y={\mbox Exp}(A)$ positive definite. $\diamondsuit$.

\begin{remark}\label{GoesThru}
{\rm The previous proof assumed that there is a basis of
one-vectors, $\{V_{i}\}$
for ${\mbox Cl}(p, q)$ with the properties {\bf BP1} and {\bf BP2}. 
For all the
Clifford algebras discussed in this paper, this is true by construction.
However, for the sake of completeness, we will prove that there is
at least one such basis for all ${\mbox Cl}(p, q)$
in Theorem \ref{SpecialB}. Notwithstanding Theorem \ref{SpecialB}, it is worth
emphasizing that for the purpose
of inversion, in light of Remark \ref{CVA}, 
one must verify the veracity of both {\bf BP1} and {\bf BP2} 
for the specific basis of one-vectors that one chooses
to arrive at the matrix form
of $\Phi_{p,q}$.}
\end{remark}

\subsection{Polar and Givens Decomposition of $SO^{+}(p, q)$}\label{Section 2.4}

In this section we collect together various results on
decompositions of $SO^{+}(p, q)$ which will play an important
role in the remainder of this work.

\begin{remark}\label{PDforn1}
{\rm Constructive Polar Decomposition:
Let $X\in SO^{+}(p, q)$. Then (see \cite{jciss}),
both factors $V, P$ in its polar decomposition $X = VP$, where $V$ is
real  special orthogonal and $P$ is positive definite,  also belong to
$SO^{+}(p, q)$. Furthermore, as shown
in \cite{jciss}, when either $p$ or $q$ is $1$
one can find $V, P$ and the real symmetric $Q$ with ${\mbox Exp}(Q) = P$
by inspecting the first column and row 
(or last if $p = 1$) and finding special orthogonal matrices which take
the first unit vector to a given vector of length one. See Algorithm
\ref{PDbyRotation} in Section \ref{Section 3} 
below for a special case of this. 
For other values of $(p, q)$ these constructive  procedures
can be extended, except
that they involve substantially more matrix maneuvers. We will
tacitly assume the contents of 
this remark in Sections \ref{Section 3} and \ref{Section 5}.
}
\end{remark}

\begin{remark}\label{FindingLog} Logarithms of Special Orthogonal Matrices of Size 4.\ \\
{\rm Let $X\in {\rm SO(4)}$. Then one can find explicitly a pair of unit
quaternions $u, v$ such $X = M_{u\otimes v}$, (see, for instance,
\cite{FourportI}). Suppose first that neither $u$ nor $v$ belong to
the set $\{\pm 1 \}$. This means $M\neq \pm I_{4}$. 

Then one can further find, essentially  by inspection of $u, v$,
a real skew-symmetric $Y$ such that ${\mbox Exp}(Y) = X$. 
Specifically, let $\lambda \in (0, \pi )$ be
such that ${\Re}(u) = \cos (\lambda )$. Then let
$p = p_{1}i + p_{2}j + p_{3}k$ be $\frac{\lambda}{\sin (\lambda )}
{\Im}(u)$. Similarly, find  $q = q_{1}i + q_{2}j + q_{3}k$ from
inspecting $v$. Then 
\[
Y = Y_{1} + Y_{2}
\]
with
\begin{equation}
\label{Y1}
Y_{1} = \left ( \begin{array}{cccc}
0 & -p_{1} & -p_{2} & -p_{3}\\
p_{1} & 0 & -p_{3} & p_{2}\\
p_{2} & p_{3} & 0 & -p_{1}\\
p_{3} & -p_{2} & p_{1} & 0
\end{array}
\right )
\end{equation}

and
\begin{equation}
\label{Y2}
Y_{2} = \left ( \begin{array}{cccc}
0 & q_{1} & q_{2} & q_{3}\\
-q_{1} & 0 & -q_{3} & q_{2}\\
-q_{2} & q_{3} & 0 & -q_{1}\\
-q_{3} & -q_{2} & q_{1} & 0
\end{array}
\right ).
\end{equation}
Furthermore, $Y_{1}$ and $Y_{2}$ commute.

Finally, if $M= I_{4}$, then $M = {\mbox Exp}(0_{4})$, while if
$M = -I_{4}$, then \ \\ $M = {\mbox Exp} (Y)$, with
\[
Y = Z\oplus Z
\]
where $Z = \left (\begin{array}{cc}
0 & \pi\\
-\pi & 0
\end{array}
\right )$.}

\end{remark} 

Next we discuss {\it Givens} decompositions.

Define
$R =\left ( \begin{array}{cc}
c & -s\\
s & c
\end{array} \right )$ where $c^{2} + s^{2} = 1$ and $H = \left (
\begin{array}{cc} 
a  & b\\
b & a
\end{array} \right )$, for $a^{2} - b^{2} = 1$. 
Then the following facts are  well known: 
\begin{itemize}
\item Given a vector $(x,y)^{T}$ there is an $R = \left (\begin{array}{cc}
c & -s\\
s & c
\end{array}\right )$ where $c^{2} + s^{2} = 1$ 
such that $R \left (\begin{array}{c}
x\\
y
\end{array}
\right ) =
\left (\begin{array}{c} 
\sqrt{x^{2}+y^{2}}\\
0
\end{array}
\right )$. Similarly there is an $R = \left (\begin{array}{cc}
c & -s\\
s & c
\end{array}\right )$ where $c^{2} + s^{2} = 1$ such that 
$R \left (\begin{array}{c}
x\\
y
\end{array}
\right ) =
\left (\begin{array}{c}
0\\ 
\sqrt{x^{2}+y^{2}}
\end{array}
\right )$.

\item Next given a vector $(x,y)^{T}$, with
$\mid x\mid \geq \mid y\mid$, there is an $H = \left (\begin{array}{cc}
a  & b\\
b & a
\end{array}\right )$ 
with $a^{2} - b^{2} = 1$, such that $H\left (\begin{array}{c}
x\\
y
\end{array}
\right ) =
\left (\begin{array}{c}
\sqrt{x^{2}-y^{2}}\\
0
\end{array}
\right ) 
$
\end{itemize}

$R, H$ are called plane standard Givens and hyperbolic Givens respectively.
Embedding $R$, respectively $H$ as a principal submatrix of the
identity matrix $I_{n}$, yields matrices known as
standard Givens (respectively, Hyperbolic Givens).

\begin{definition}
{\rm $H_{ij}$, for $i < j$, stands 
for the $n\times n$ matrix which is the identity
except in the principal submatrix, indexed by rows and columns $(i,j)$,
wherein it is a hyperbolic Givens matrix. Similarly,
$R_{ij}$ stands for the $n\times n$ matrix which is the identity
except in the principal submatrix, indexed by rows and columns $(i,j)$,
wherein it is an ordinary Givens matrix.}
\end{definition}

\begin{remark}
\label{Caution}
{\rm While $H_{ij}$ is defined only if $i < j$, the matrices $R_{ij}$
make sense for all pairs $(i, j)$ with $i\neq j$. $R_{ij}$
is the matrix which zeroes out the $j$th component of a vector that
it premultiplies. Thus, $R_{ij}$ will be different, in general, if
$i < j$ from that when $i > j$.}
\end{remark}

The next result shows that $H_{ij}$'s and $R_{ij}$'s
belong to $SO^{+}(p, q)$. Specifically
\begin{proposition}
\label{inrightplace}
{\rm Let $1\leq i \leq p$ and $1\leq j \leq q$. Then $H_{ij}$ belongs
to $SO^{+}(p, q)$. Similarly, if either $1\leq i, j\leq p$ or
$1\leq i, j\leq q$, then $R_{ij}$ belongs
to $SO^{+}(p, q)$.}
\end{proposition}

\noindent {\bf Proof:} Consider the $H_{ij}$ case first. Define
\[
L_{ij} = \theta (e_{i}e_{j}^{T} + e_{j}e_{i}^{T})
\]
with $\theta \in \mathbf{R}$.
Thus $L_{ij}$ is the symmetric matrix which is zero everywhere, except
in the $(i,j)$th and $(j,i)$th entries wherein it is $\theta$.
Due to the conditions, $1\leq i \leq p$ and $1\leq j \leq q$, it is
easy to verify that $L_{ij}\in so(p, q)$. A calculation shows that
\[
L_{ij}^{2} = D_{\theta},
\]
where $D_{\theta}$ is $n\times n$ diagonal with zeroes everywhere,
except on the $i$th and $j$th diagonal entries wherein it is $\theta^{2}$.  
Therefore,
\[
L_{ij}^{3} = \theta^{3} (e_{i}e_{j}^{T} + e_{j}e_{i}^{T})
= \theta^{2}L_{ij}.
\]

Hence by the Euler-Rodrigues formula
\[
{\mbox Exp} (L_{ij}) = I_{n} + \frac{\sinh (\theta )}{\theta}L_{ij} +
\frac{\cosh (\theta ) -1}{\theta^{2}} L_{ij}^{2}
\]
\[ 
= I_{n} + \sinh (\theta )(e_{i}e_{j}^{T} + e_{j}e_{i}^{T}) +
\frac{\cosh (\theta ) -1}{\theta^{2}}D_{\theta}\\
= H_{ij}.
\]

Thus $H_{ij}$ being the exponential of a matrix in the Lie algebra
$so(p, q)$, belongs to $SO^{+}(p, q)$.
The proof for $R_{ij}$ is similar. $\diamondsuit$ 

The relevance of Givens rotations is that any matrix
in ${\rm SO^{+}}(p, q)$ can be decomposed constructively
into a product of Givens matrices. It will suffice to
illustrate this via an example.

\begin{example}

\label{Givens}
{\rm Let $X\in {\rm SO^{+}}(2,2)$. Consider the first column of $X$

\[
v_{1} = \left ( \begin{array}{c}
a \\ b\\ c\\ d
\end{array}
\right ).
\]
Since $X\in {\rm SO^{+}}(2,2)$, $a^{2}+b^{2} -c^{2}- d^{2} = 1$.
Therefore there are  $R_{1,2},R_{3,4}$ such that the first column of
$R_{1,2},R_{3,4}X =  
\left ( \begin{array}{c} 
\alpha \\ 0 \\ \beta \\ 0
\end{array}
\right )$,
where $\alpha = \sqrt{a^{2}+b^{2}}$ and $\beta = \sqrt{c^{2}+d^{2}}$.
 Since  $a^{2}+b^{2} -c^{2}- d^{2} = 1 = \alpha^{2} - \beta^{2}$,
 it follows that $\vert \alpha \vert > \vert \beta \vert$. Hence there is an
 $H_{1,3}$ such that the first column of 
$H_{1,3}R_{1,2}R_{3,4}X$ equals 
\[
\left ( \begin{array}{c}
1\\ 0\\ 0\\ 0
\end{array} \right ). 
\]

As $H_{1,3}R_{1,2}R_{3,4}X \in {\rm SO}^{+}(2,2)$, it follows that 
the first row of $H_{1,3}R_{1,2}R_{3,4}X$ is also $
  ( 1,   0,   0,    0 )$.

 Therefore, the second column of the product 
$H_{1,3}R_{1,2}R_{3,4}X$ is of the form
 $\left ( \begin{array}{c}
 0 \\ b \\ c \\ d
 \end{array} \right )$
 with $b^{2} -c^{2}- d^{2} = 1$. 
So there is an $R_{3,4}$ such that
 $R_{3,4}(c,d)^{T} = (\gamma , 0)^{T} $, 
where $\gamma^{2} =c^{2}+ d^{2}$. As before $b^{2}-\gamma^{2} = 1$, 
so there is an $H_{2,3}$ such that $H_{2,3}$
with $H_{2,3}(b,\gamma )^{T} = (1,0)$. 
So it follows that the first and the second column equal the 
first two standard unit vectors.
Since $H_{2,3}R_{3,4}H_{1,3}R_{1,2}R_{3,4}X \in {\rm SO^{+}}(2,2)$,
it follows that it equals\\
	\[
     \left (\begin{array}{cccc}
     1 & 0 & 0 & 0\\
     0 & 1 & 0 & 0\\
     0 & 0 & y_{33} & y_{34}\\
     0 & 0 &  y_{43} & y_{44}
\end{array}
\right )
\]

 Again the condition 
$H_{2,3}R_{3,4}H_{1,3}R_{1,2}R_{3,4}X \in {\rm SO^{+}}(2,2)$, 
ensures that
 \[ 
     \left (\begin{array}{cc}
      y_{33} & y_{34}\\
      y_{43} & y_{44}
     \end{array} \right )
\]
 must itself be a plane standard Givens rotation.
Therefore, pre-multiplying by the corresponding
$R_{3,4}$ we get that
\[
 R_{3,4}H_{2,3}R_{3,4}H_{1,3}R_{1,2}R_{3,4}X = I_{4}.
\]

 Since the inverse of each $R_{i,j}$ (respectively $H_{k,l}$) is itself an
 $R_{i,j}$ (respectively $H_{k,l}$), it follows that $X$ can be expressed
 {\it constructively} as a product of
 $R_{3,4}$, $H_{2,3}$, $H_{1,3}$, $R_{1,2}$.
}
\end{example}
\begin{remark}\label{However}
{\rm
The following observations about Givens decomposition are pertinent 
for this work.
\begin{itemize}
\item[i)] The above factorization is not the only way to factor an element of
${\rm SO^{+}}(2,2)$ into a product of standard and hyperbolic Givens matrices.
By way of illustration, we use a slightly different factorization in
Section \ref{Section 4}, which emanates from using an $R_{4,3}$ instead of an $R_{3,4}$
in one of the three usages of $R_{3,4}$ above. This will result,
therefore, in the usage of an $H_{2,4}$ instead of an $H_{2,3}$.
We note, however, that since $R_{3,4}$ and an $R_{4,3}$ are essentially
the same matrix, differing only in the parameter $\theta$ which enters
in them. Thus, their inversion will require the symbolic solution of
the same system of equations.
\item[ii)] There are atmost $\left(\begin{array}{c}
p+q\\
2
\end{array}\right)$ Givens factors 
in the decomposition of a generic $X$. However,
of these there are at most $2p + q -2$ distinct such factors. This is
pertinent, as it implies that we have to 
symbolically invert only $2p + q -2$ targets.
\end{itemize}
}
\end{remark}

\section{Inversion of $\Phi_{2,1}$ and the Polar Decomposition}\label{Section 3} 
In this section, we treat the inversion of $\Phi_{2,1}:
{\mbox Spin}^{+}(2,1) \rightarrow {\mbox SO^{+}}(2,1)$ by showing that
it only requires inspection to find the preimage (under $\Phi_{2,1}$)
of matrices in ${\rm SO^{+}}(2,1)$ which are either positive definite
or special orthogonal. Since the factors in the polar decomposition
of an $X\in {\rm SO^{+}}(2,1)$ also belong to ${\rm SO^{+}}(2,1)$, the method
below also simultaneously provides the polar decomposition of the
$X$ being inverted, with minimal fuss. Alternatively, one can also
directly find the polar decomposition of $X$,
by essentially inspecting the last row and some extra calculations,
and use that to invert
$\Phi_{2,1}$. 

In principle these methods
extend to all $(p, q)$ but are limited in that, besides the
$(2,1)$ and $(3,1)$ cases (the latter is treated in \cite{jciss}),
finding the preimage by mere inspection seems difficult. 
See however,
Section \ref{Section 4}, wherein the $(4,1)$ case is handled by a combination
of the polar decomposition and inverting the associated Lie algebra
isomorphism $\Psi_{4,1}: {\mbox spin}^{+}(4,1) \rightarrow
{\mbox so}(4, 1)$.

Let us first provide an explicit matrix form of the map $\Phi_{2,1}$. 
This follows, after some computations,
from  the material in \cite{fpi}.
Specifically we begin with the following
basis of one-vectors for ${\mbox Cl}(2,1)$

\begin{equation}
\label{Basis21}
{\mathbf B}_{2,1} = \{Y_{1}, Y_{2}, Y_{3}\}
= \{ \sigma_{z}\otimes \sigma_{z}, 
 \sigma_{x}\otimes I_{2},
i\sigma_{y}\otimes I_{2}\}.
\end{equation}

Thus, ${\mbox Cl}(2,1)$ is a matrix subalgebra
of $M(4, \mathbf{R})$.
The even subalgebra is isomorphic to $M(2, \mathbf{R})$, 
which can be embedded
into the former subalgebra as follows.  
Specifically, given $Y = \left (\begin{array}{cc}
y_{1} & y_{2}\\
y_{3} & y_{4}
\end{array}
\right ) \in {\mbox Cl}(2,1)$ embed it in ${\mbox Cl}(2,2)$ as follows
\begin{equation}\label{Embed21}
\left (\begin{array}{cccc}
y_{1} & 0 & y_{2} & 0\\
0 & -y_{1} & 0 & y_{2}\\
y_{3} & 0 & y_{4} & 0\\
0 & y_{3} & 0 & -y_{4}
\end{array}
\right ).
\end{equation}

Then ${\mbox Spin^{+}}(2,1)$ is isomorphic to ${\rm SL}(2, \mathbf{R})$
(embedded in $M(4, \mathbf{R})$ as in Equation (\ref{Embed21})  above).
 
It can then be shown that the map $\Phi_{2,1}$ sends an
element 
$\left (\begin{array}{cc}
y_{1} & y_{2}\\
y_{3} & y_{4}
\end{array}
\right )$ in ${\rm SL}(2, \mathbf{R})$ to the following matrix in
${\rm SO^{+}}(2,1)$
\begin{equation}\label{ImPhi21}
\left( 
\begin{array}{ccc}
1+2y_{2}y_{3} & y_{2}y_{4}-y_{1}y_{3} & -\left( y_{1}y_{3}+y_{2}y_{4}\right)
\\ 
y_{3}y_{4}-y_{1}y_{2} & \frac{1}{2}\left(
y_{1}^{2}-y_{2}^{2}-y_{3}^{2}+y_{4}^{2}\right) & \frac{1}{2}\left(
y_{1}^{2}+y_{2}^{2}-y_{3}^{2}-y_{4}^{2}\right) \\ 
-\left( y_{1}y_{2}+y_{3}y_{4}\right) & \frac{1}{2}\left(
y_{1}^{2}-y_{2}^{2}+y_{3}^{2}-y_{4}^{2}\right) & \frac{1}{2}\left(
y_{1}^{2}+y_{2}^{2}+y_{3}^{2}+y_{4}^{2}\right)%
\end{array}
\right).
\end{equation}

\subsection{Preimages of Positive Definite Targets in ${\rm SO^{+}}(2,1)$}\label{Section 3.1}
First from many points of view, e.g., from Equation (\ref{ImPhi21}) itself,
it is easily seen that if $Y\in {\rm SL}(2,\mathbf{R})$, then
$\Phi_{2,1}(Y^{T}) = [\Phi_{2,1}(Y)]^{T}$. Hence, in view of the fact that $\Phi_{2,1}$ is surjective
with ${\ker}(\Phi_{2,1}) = \{ \pm I\}$,
we see that if $\Phi_{2,1}(Y)$
is a symmetric matrix in ${\rm SO^{+}}(2,1)$ then necessarily $Y^{T} = \pm Y$.

Next, a symmetric $Y$  in ${\rm SL}(2, \mathbf{R})$ cannot have its 
$(1,1)$ entry
equal to zero. Thus, as ${\det}(Y) > 0$, $Y$
is either positive or negative definite. If $Y$ is antisymmetric,
it is easily seen from Equation (\ref{ImPhi21})
that $\Phi_{2,1}(Y)$ is diagonal with two entries
negative and one positive - i.e., it is indefinite.

Furthermore,
if $Y\in {\rm SL}(2, \mathbf{R})$ is symmetric
then one can also deduce directly that $\Phi_{2,1}(Y)$ is positive definite.
To that end, note that since ${\det} (\Phi_{2,1}(Y)) = 1$
and quite visibly
the $(1,1)$ entry is positive, it suffices
to check that the $(1, 2)$ minor is positive to verify positive definiteness. 

From Equation (\ref{ImPhi21}), this minor equals
\[
\frac{1}{2} (1 + 2y_{2}^{2}) (y_{1}^{2} - 2y_{2}^{2} + y_{4}^{2})
- y_{2}^{2} (y_{1} - y_{4})^{2}.
\]
Using $y_{1}y_{4} - y_{2}^{2} = 1$ we find that this minor is
$\frac{1}{2}(1 + 2y_{2}^{2})[ (y_{1} - y_{4})^{2} + 2]
- y_{2}^{2}(y_{1}-y_{4})^{2} = 2y_{2}^{2} + (1/2)(y_{1}- y_{4})^{2}
+ 1 > 0$.

Summarizing the contents of the previous two paragraphs we have
\begin{theorem}\label{PosDefin21}
{\rm Let $X\in {\rm SO^{+}}(2,1)$ be symmetric. Then it is either positive
definite or indefinite. In the former case $X = \Phi_{2,1}(\pm Y)$,
with $Y\in {\rm SL}(2, \mathbf{R})$ also positive definite. In the latter
case $X$ is diagonal and $X =  \Phi_{2,1}(\pm Y)$ with $Y\in
{\rm SL}(2, \mathbf{R})$ antisymmetric.} 
\end{theorem}

\subsection{Finding $\Phi_{2,1}^{-1}(X)$ when $X> 0$ by Inspection}\label{Section 3.2}
Suppose that $X$ is positive definite. Let us then address how a positive definite preimage $Y\in {\rm SL}(2, \mathbf{R})$ is 
found by inspection of Equation (\ref{ImPhi21}). Let $Y =
\left (\begin{array}{cc}
y_{1} & y_{2}\\
y_{2} & y_{4}
\end{array}
\right )$.
From the $(1,1)$ entry of Equation (\ref{ImPhi21}) 
we see $y_{2} = \pm \frac{1}{\sqrt{2}}\sqrt{X_{11} -1}$. 
Suppose $X_{11}\neq 1$ first.
Then we find $y_{1}, y_{4}$ from the equations
\[
y_{4} - y_{1} =  \frac{X_{12}}{y_{2}}, \quad y_{4} + y_{1}  =  -\frac{X_{13}}{y_{2}}\cdot
\]
By Theorem \ref{PosDefin21},
one choice of the sign for $y_{2}$ will lead to a $Y$ which is positive
definite. If $X_{11} = 1$, then $y_{2} = 0$. Now we look at $X_{22}$
and $X_{23}$ to find that $y_{1}^{2}$ and $y_{4}^{2}$ may be found by
solving the system
\[
y_{1}^{2} + y_{4}^{2}  =  2X_{22}, \quad y_{1}^{2} - y_{4}^{2} = 2X_{23}.
\]
We take the positive square roots of the solutions $y_{1}^{2}$ and $y_{4}^{2}$
to find a positive definite $Y$ projecting to $X$.
This finishes our claim that if $X\in {\rm SO^{+}}(2,1)$ is positive definite
then we can find by inspection a positive definite
$Y\in {\rm SL}(2, \mathbf{R})$ projecting to $X$ under $\Phi_{2,1}$  

The above discussion is 
summarized in Algorithm \ref{Phi21InverseofPosDef} below.

\begin{algorithm}\label{Phi21InverseofPosDef}
{\rm Let $X = (X_{ij})\in {\rm SO^{+}}(2,1)$ be positive definite. 
The following algorithm
finds a positive definite $Y\in {\rm SL}(2, \mathbf{R})$ satisfying
$\Phi_{2,1}(Y) = X$.
\begin{itemize} 
	
\item[1.] Suppose $X_{11}\neq 1$. Let $y_{2} = 
\pm \frac{1}{\sqrt{2}}\sqrt{X_{11} -1}$, $y_{1} = -\frac{\displaystyle X_{12} + X_{13}}{\displaystyle 2y_{2}}$ and $y_{4} =\frac{\displaystyle X_{12} - X_{13}}{\displaystyle 2y_{2}}\cdot$ 

\item[2.] Let $Y = \left (\begin{array}{cc}
y_{1} & y_{2}\\
y_{2} & y_{4}
\end{array}
\right )$. There are two choices of $Y$ corresponding to the choice of the square root in $y_{2}$ in Step 1, which are negatives of each other. One of these $Y$ is positive definite. Pick this one.

\item[3.] Suppose $X_{11} = 1$. Then let $y_{2} = 0$, $y_{1}
= \sqrt{X_{22} + X_{23}}$ \enspace and $y_{4} = \sqrt{X_{22} - X_{23}}$.
Then $Y = {\mbox diag}(y_{1}, y_{4})$ is positive definite and is 
one preimage of $X$ in ${\rm SL}(2, \mathbf{R})$.
\end{itemize}
}
\end{algorithm}
  
\subsection{Finding the polar decomposition in ${\rm SO^{+}}(2,1)$}\label{Section 3.3}
 
Let us now address how to find the polar decomposition in ${\rm SO^{+}}(2,1)$
using Algorithm \ref{Phi21InverseofPosDef}.

Let $X = VP$ be the polar decomposition of an $X\in {\rm SO^{+}}(2,1)$.
Then the orthogonal $V$ and positive definite $P$ are both in
${\rm SO^{+}}(2,1)$. To find $P$ we proceed as follows. First find 
$X^{T}X$, which is then a positive definite element of
${\rm SO^{+}}(2,1)$. Since
its preimage $Z$ can be chosen to be positive definite, we find it
using Algorithm \ref{Phi21InverseofPosDef}.  
Once $Z$ has been found, we compute its unique positive
definite square root, $W$. We note in passing that finding $Y$ from
$Z$ can be executed in closed form, without any eigencalculations,
\cite{jciss}. 
 
Since $Z\in SL(2, \mathbf{R})$, $W$ is also in $SL(2, \mathbf{R})$.
Then let $P = \Phi_{2,1}(W)\in {\rm SO^{+}}(2,1)$.
Then, we compute
\[
P^{2} = P^{T}P = [\Phi_{2,1}(Y)]^{T}\Phi_{2,1}(Y) = 
\Phi_{2,1}(Y^{T})\Phi_{2,1}(Y)
\]
\[
= \Phi_{2,1}(Y^{T}Y) = \Phi_{2,1}(Z)= X^{T}X.
\]
So $P$ is the positive definite factor
in $X= VP$. Of course, $V = XP^{-1}$. Next,
finding $P^{-1}$ is easy. One just
interchanges $y_{1}$ and $y_{4}$ and replaces $y_{2}$ by $-y_{2}$ and
$y_{3}$ by $-y_{3}$ in the formula $\Phi_{2,1}(Y) = P$. This completes
the determination of the polar decomposition of $X$.

However, for the purpose of
inversion of $\Phi_{2,1}$,
it still remains to find
$ \pm S\in SL(2, \mathbf{R})$ satisfying $\Phi_{2,1}(\pm S) = V$.  

To that end, note first that $V$ is both orthogonal and in ${\rm SO^{+}}(2,1)$. 
Thus, it is in
${\rm SO}(3)$. Hence it
must have the following form $V = {\mbox diag}(R, \pm 1)$, where
$R$ is $2\times 2$ orthogonal. 
However, from Equation (\ref{ImPhi21})
it is clear that the $(3,3)$ entry 
of a matrix in ${\rm SO^{+}}(2,1)$ is positive.
So, $V = {\mbox diag}(R, 1)$, with $R$ in $SO(2)$. 
Let $c=\cos \theta$, $s=\sin \theta $. Then
$$V ={\left( 
\begin{array}{rrr}
c & -s & 0 \\ 
s & c & 0 \\ 
0 & 0 & 1%
\end{array}%
\right).}$$

As before, simple considerations show that the matrix $S\in 
SL(2, \mathbf{R})$
projecting to $R$ must itself be in ${\rm SO}(2)$.
Finding $R$'s entries as functions of $c$ and $s$ is easy. First, if $\theta \in \left(
0,2\pi \right)$, then $\sin \left( \frac{\theta }{2}\right) >0$.
Indeed, denoting by
$\widehat{c}=\cos \left( \frac{\theta }{2}\right)$, 
$\widehat{s}=\sin \left( 
\frac{\theta }{2}\right)$, 
we have
\begin{equation}
\label{InvertOrthoin211}
S =\pm \left( 
\begin{array}{rr}
\widehat{c} & -\widehat{s} \\ 
\widehat{s} & \widehat{c}  
\end{array}%
\right).
\end{equation}
For $\theta =0,2\pi $ we get 
\begin{equation}
\label{InvertOrthoin212}
S = \pm {\left( 
\begin{array}{rr}
-1 & 0 \\ 
0 & -1  
\end{array}%
\right) }.
\end{equation}
We summarize all of this in an algorithm
\begin{algorithm}\label{PDfor21ViaPhi21}
{\rm Given $X\in SO^{+}(2,1)$, the following algorithm
computes both its polar decomposition and the $Y\in SL(2, \mathbf{R})$
satisfying $\Phi_{2,1}(\pm Y) = X$.
\begin{itemize}
\item[1.] Find $X^{T}X$ and find $Z\in SL(2, \mathbf{R})$ positive
definite such that $\Phi_{2,1}(Z) = X^{T}X$, using Algorithm
\ref{Phi21InverseofPosDef}.
\item[2.] Find the unique positive definite square root $W$
of $Z$. This step can be executed without any diagonalization
- see \cite{jciss}.
\item[3.] Find $P = \Phi_{2,1}(W)$ using Equation (\ref{ImPhi21}).
\item[4.] Find $P^{-1}$ by interchanging $p_{11}$ and $p_{22}$ and
replacing $p_{12}, p_{21}$ by $-p_{12}, -p_{21}$ 
respectively in $P$ from Step 3.
\item[5.] Find $V=XP^{-1}$. Then $X=VP$ is the polar decomposition
of $X$.
\item[6.] Find $S\in {\rm SO}(2)$ satisfying 
$\Phi_{2,1}(S) = V$ using Equation (\ref{InvertOrthoin211})
or Equation (\ref{InvertOrthoin212}). Then $Y = WS$ satisfies
$\Phi_{2,1}(\pm Y) = X$.
\end{itemize}
}
\end{algorithm}
The above algorithm inverts the covering map by also finding the polar decomposition of the matrix in ${\rm SO^{+}}(2,1)$. However, we can also find the polar decomposition without using the covering map.
We now present a  second algorithm which will produce the polar decomposition
directly from $X$ itself, with inspection and finding special orthogonal
matrices which rotate a plane vector into another of the same length.
This is a special case of the algorithm for general ${\rm SO}^{+}(n,1)$
from \cite{jciss} mentioned earlier in Remark \ref{PDforn1}.

\begin{algorithm}\label{PDbyRotation}
The Polar Decomposition from the last row and column of $X$
{\rm \begin{itemize} 
\item[1.] Define $\sigma> 0$ by $X_{33} = \cosh (\sigma )$. If $X_{33}= 1$, then $X=I_{3}$ and the polar decomposition of $X$ is trivial.
\item[2.] Find $U\in {\rm SO}(2)$ so that $(X_{31}, X_{32})
= (\sinh (\sigma ), 0)U^{T}$.
\item[3.] Find $Z\in {\rm SO}(2)$ so that $ZU\left (\begin{array}{c}
\sinh (\sigma )\\
0
\end{array}
\right ) =  \left (\begin{array}{c}
X_{13}\\
X_{23}
\end{array}
\right )$
		
\item[4.] Define i) $C_{1\times 1} = (\cosh (\sigma ))$, ii)
$\tilde{C} = C\oplus I_{1}$ and iii) $S_{2\times 1} =
\left (\begin{array}{c}
\sinh (\sigma )\\
0
\end{array} \right )$. Then the polar decomposition of $X$ is $X = V P$, with
$V = Z\oplus I_{1}$ and 
\[P =
\left (\begin{array}{ll}
U\tilde{C}U^{T} & US\\
S^{T}U^{T} & \cosh (\sigma )
\end{array}
\right )\]
		
\item[5.] Finally the symmetric $Q\in so(2,1)$ satisfying ${\mbox Exp} (Q) 
= P$ is
\[Q = \left (\begin{array}{ll}
0_{2\times 2} & UE\\
E^{T}U^{T} & 0_{1\times 1}
\end{array}
\right )\] with $E = \left (\begin{array}{c}
\sigma  \\
0
\end{array}
\right )
$. 
\end{itemize}
}
\end{algorithm}

\section{Inversion of $\Phi_{2,2}$}\label{Section 4}

In this and the next section, we will use Groebner bases to invert 
$\Phi_{2,2}$ and $\Phi_{2,3}$. We begin with a few observations,
which are pertinent to this and the next section, collected as remarks
for ease of future reference. 

\begin{remark}
\label{Outline}
{\rm The task of inversion of the covering maps in question, will
be addressed by solving directly the system of equations 
$\Phi_{p, q}(Y) = G$, where $G$ is either a standard Givens matrix
$G(c, s)$ or a hyperbolic Givens $H(a, b)$.
In the former case $c^{2} + s^{2} = 1$ and in the 
latter $a^{2} - b^{2} = 1$. For each fixed $c, s$ (respectively $a, b$)
this is a system of $n^{2}$ equations in the variables $y_{1}, \ldots , y_{l}$,
where $n= p + q$ and the number $l$ is determined by the structure of the
corresponding spin group. For instance, for $(p, q) = (2,2)$, $l$ is $8$,
while for $(p, q) = (3,2)$, $l$ is $16$. 
Furthermore, these equations have degree at
most one in $c$ and $s$ (respectively $a$ and $b$), thereby ensuring that the
dependence of these equations in $c$ and $s$ (respectively $a$ and $b$) is
rational $c$ and $s$ (respectively $a$ and $b$). 
Next, this system of equations has
precisely two solutions. This is guaranteed by Proposition \ref{inrightplace}, since
that result confirms that such Givens matrices reside in ${\rm SO^{+}}(p, q)$.
To take advantage of this finiteness of the solution set, we view the 
associated ideal as an ideal in the polynomial ring $K[x_{1}, \ldots , x_{n}]$,
where $K$ is the field of rational functions, with complex coefficients,
in the variables $c$ and $s$ (respectively $a$ and $b$). 
Then this ideal becomes zero dimensional and the standard wisdom suggests that we should find Groebner bases with
respect to some lexicographic order in the variables $y_{i}$ (in the polynomial
ring $K[y_{1}, \ldots , y_{l}]$). To this Groebner basis, we finally
append formally the equations $c = \cos (\theta ); s =
\sin (\theta )$ (repectively,
$a = \cosh (\beta ); b = \sinh (\beta )$). This last output is then
used to solve for the $y_{j}$'s parametrically in $\theta$ (respectively $\beta$).
It is noted in passing that 
the equations $\Phi_{p, q}(Y) = X$ are all quadratic (and most of them
actually homogeneous quadratic) and there
are techniques, besides Groebner bases (such as those related to Sylvester
matrices), for analyzing such equations.}
\end{remark}

\begin{remark}\label{NotOnly}
{\rm It is worth pointing out that Groebner bases play a twin role in the
methods of this and the next section. The first has been already outlined
above. The second, equally important, is to arrive at the matrix form of
the covering maps $\Phi_{p, q}$. Indeed, the entries of the matrix
representing $\Phi_{p, q}(Y)$ were arrived at by computing the matrix of
the linear map $V \rightarrow YVY^{-1}$, where $Y$ is an element of
the spin group and $V$ is a one-vector, with respect to the basis
of one-vectors $\{V_{i}\}$ chosen for this purpose. So to compute $\Phi_{p, q}(Y)$,  we compute the traces of $YV_{i}Y^{-1}V_{j}^{*}$ modulo the defining
relations for the spin group. In all instances, these defining relations
are some quadratic relations in the entries of the typical element of
the matrix algebra that ${\mbox Cl}(p, q)$ is. For instance, for $(p, q) = (2,2)$,
these relations are two equations representing the fact that the determinants
of a pair of matrices, representing the typical element of
${\mbox Spin^{+}}(2,2)$ both equal $1$.}
\end{remark}

Now let us turn to the inversion of $\Phi_{2,2}$.
    
Let $X\in {\rm SO^{+}}( 2, 2) $. First, following Example \ref{Givens} and with the caveat in Remark \ref{However}, it is noted that $X$ can be represented as 
a product of ordinary and hyperbolic Givens rotations (non-uniquely)
as follows
\begin{equation}
\label{Givensfor22}
X = R_{1,2}R_{3, 4}H_{1,3}R_{4,3}H_{2,4}R_{3,4}.
\end{equation} 
So it suffices to find the preimages of these Givens matrices
in $SL(2, \mathbf{R})\times SL(2, \mathbf{R})$, the relevant spin group. 
As mentioned in Remark \ref{However}, the systems of equations for
the inversion of $R_{3,4}$ and $R_{4,3}$ are essentially the same.  Furthermore by ii) of Remark \ref{However} we have to only solve four such systems symbolically.

To get to that goal, we begin first with an explicit description of
the entries of the matrix $\Phi_{2,2}(Y), Y\in SL(2, \mathbf{R})\times 
SL(2, \mathbf{R})$.

Following \cite{fpi}, 
the following quartet of matrices is used for a basis of one-vectors
for ${\mbox Cl}(2,2)$   

\begin{equation}\label{Basis22} 
{\mathbf B}_{2,2} = \{ X_{1}=\sigma _{z}\otimes \sigma _{x}, 
X_{2}=\sigma _{x}\otimes I, 
X_{3}=\sigma _{z}\otimes (i\sigma _{y}),  
X_{4}= (i\sigma_{y}) \otimes I \}.
\end{equation}

Again, as described in \cite{fpi}, we can view a pair
of matrices $$\{ \left (\begin{array}{cc}
y_{1} &  y_{2} \\
y_{7} &  y_{8}
\end{array}
\right ), \left (\begin{array}{cc}
y_{3} & y_{4}\\ 
y_{5} & y_{6}
\end{array}
\right )\}$$ as embedded in ${\mbox Cl} (2,2) = M(4, \mathbf{R})$ as follows 
\begin{equation}
\label{Embed22} 
Y
=\left( 
\begin{array}{cccc}
y_{1} & 0 & 0 & y_{2} \\ 
0 & y_{3} & y_{4} & 0 \\ 
0 & y_{5} & y_{6} & 0 \\ 
y_{7} & 0 & 0 & y_{8}%
\end{array}%
\right). 
\end{equation}
 
The variables $y_{1}, \ldots , y_{8}$ satisfy the quadratic relations
\begin{equation}\label{22Relations}
y_{1}y_{4} - y_{2}y_{3} - 1 = 0,\quad y_{5}y_{8}-y_{6}y_{7} -1 = 0.
\end{equation}
The entries of the matrix
$\Phi_{2,2}(Y)$ are then found by calcuating the matrix of the 
linear map $V\rightarrow YVY^{-1}$,
with $Y$ as in Equation (\ref{Embed22}), with respect to the basis
of one-vectors in Equation (\ref{Basis22}), modulo the relations
in Equation (\ref{22Relations}) (cf., Remark \ref{NotOnly}.  
Then a Groebner basis aided calculation shows that $\Phi_{2,2}$ sends $Y$
to \ \\ $D\in {\rm SO}^{+}(2,2)$, where the entries of $D$ are given by
\begin{small}
\begin{eqnarray*}
&&
\begin{array}{ccc}
d_{11} & = & \frac{1}{2}\left(
y_{2}y_{5}+y_{1}y_{6}+y_{4}y_{7}+y_{3}y_{8}\right)  \\ 
d_{21} & = & \frac{1}{2}\left(
y_{6}y_{7}+y_{5}y_{8}-y_{2}y_{3}-y_{1}y_{4}\right)  \\ 
d_{31} & = & \frac{1}{2}\left(
y_{2}y_{5}+y_{1}y_{6}-y_{4}y_{7}-y_{3}y_{8}\right)  \\ 
d_{41} & = & -\frac{1}{2}\left(
y_{2}y_{3}+y_{1}y_{4}+y_{6}y_{7}+y_{5}y_{8}\right) 
\end{array}
\begin{array}{ccc}
d_{12} & = & \frac{1}{2}\left(
y_{2}y_{6}-y_{1}y_{5}-y_{3}y_{7}+y_{4}y_{8}\right)  \\ 
d_{22} & = & \frac{1}{2}\left(
y_{1}y_{3}-y_{2}y_{4}-y_{5}y_{7}+y_{6}y_{8}\right)  \\ 
d_{32} & = & \frac{1}{2}\left(
y_{2}y_{6}-y_{1}y_{5}+y_{3}y_{7}-y_{4}y_{8}\right)  \\ 
d_{42} & = & \frac{1}{2}\left(
y_{1}y_{3}-y_{2}y_{4}+y_{5}y_{7}-y_{6}y_{8}\right) 
\end{array}
\\
&&
\begin{array}{ccc}
d_{13} & = & \frac{1}{2}\left(
y_{1}y_{6}-y_{2}y_{5}+y_{4}y_{7}-y_{3}y_{8}\right)  \\ 
d_{23} & = & \frac{1}{2}\left(
y_{2}y_{3}-y_{1}y_{4}+y_{6}y_{7}-y_{5}y_{8}\right)  \\ 
d_{33} & = & \frac{1}{2}\left(
y_{1}y_{6}-y_{2}y_{5}-y_{4}y_{7}+y_{3}y_{8}\right)  \\ 
d_{43} & = & \frac{1}{2}\left(
y_{2}y_{3}-y_{1}y_{4}-y_{6}y_{7}+y_{5}y_{8}\right) 
\end{array}\hspace{0.11in}
\begin{array}{ccc}
d_{14} & = & -\frac{1}{2}\left(
y_{1}y_{5}+y_{2}y_{6}+y_{3}y_{7}+y_{4}y_{8}\right)  \\ 
d_{24} & = & \frac{1}{2}\left(
y_{1}y_{3}+y_{2}y_{4}-y_{5}y_{7}-y_{6}y_{8}\right)  \\ 
d_{34} & = & \frac{1}{2}\left(
y_{3}y_{7}+y_{4}y_{8}-y_{1}y_{5}-y_{2}y_{6}\right)  \\ 
d_{44} & = & \frac{1}{2}\left(
y_{1}y_{3}+y_{2}y_{4}+y_{5}y_{7}+y_{6}y_{8}\right) 
\end{array}
\end{eqnarray*}
\end{small}

To invert $\Phi_{2,2}$ we equate $D$ to each distinct $R_{ij}(c,s)$ or
$H_{kl}(a, b)$ which appears in Equation (\ref{Givensfor22}).
and solve the corresponding
system of quadratic equations in the variables $y_{1}, \ldots, y_{8}$.
The $R_{ij}$'s depend on entries $c$ and $s$, while the $H_{ij}$'s depend
on entries $a$ and $b$, where
$c=\cos \theta $, $s=\sin \theta $, $a=\cosh \left( \theta \right) $ and 
$b=\sinh \left( \theta \right)$. 

The following result records $\Phi_{2,2}^{-1}(D)$, where $D$ is one of
the Givens matrices in Equation (\ref{Givensfor22}). 
For brevity we list the 
matrices as living in $M(4, \mathbf{R})$. The corresponding
pair in $SL(2, \mathbf{R})\times SL(2, \mathbf{R})$ can easily be read
off by inspection by invoking Equation (\ref{Embed22}).

The results are then given by
\begin{theorem}\label{Invert22}
{\rm Let $c = \cos \theta$; $s = \sin \theta$, $a = \cosh\beta$, $b= \sinh \beta$.
Correspondingly, let $\widehat{c} = \cos \frac{\theta}{2}$,
$\widehat{s} = \sin \frac{\theta}{2}$.
Then, the  following list provides $Y\in SL(2, \mathbf{R})\times 
SL(2, \mathbf{R})$, viewed as
matrices in $M(4, \mathbf{R})$ as embedded by Equation (\ref{Embed22}), such that
$\Phi_{2,2}(\pm Y) = D$, where $D\in G$, with $G$ given by Equation
(\ref{Givensfor22})

\begin{tabular}{|l|l|l|l|}
	\hline
	$D=R_{1,2}$ & $\theta \in \left( 0,\pi \right) $ & $\theta \in \left( \pi
	,2\pi \right) $ & $\theta =\pi $ \\ \hline
	$\Phi _{2,2}^{-1}(D)$ & $\mp \left( 
	\begin{array}{cccc}
	\widehat{c} & 0 & 0 & -\widehat{s} \\ 
	0 & \widehat{c} & -\widehat{s} & 0 \\ 
	0 & \widehat{s} & \widehat{c} & 0 \\ 
	\widehat{s} & 0 & 0 & \widehat{c}%
	\end{array}%
	\right) $ & $\pm \left( 
	\begin{array}{cccc}
	\widehat{c} & 0 & 0 & -\widehat{s} \\ 
	0 & \widehat{c} & -\widehat{s} & 0 \\ 
	0 & \widehat{s} & \widehat{c} & 0 \\ 
	\widehat{s} & 0 & 0 & \widehat{c}%
	\end{array}%
	\right) $ & $\pm \left( 
	\begin{array}{cccc}
	0 & 0 & 0 & 1 \\ 
	0 & 0 & 1 & 0 \\ 
	0 & -1 & 0 & 0 \\ 
	-1 & 0 & 0 & 0%
	\end{array}%
	\right) $ \\ \hline
\end{tabular}%
\smallskip 

For $\theta =0,2\pi ,$ $\Phi _{2,2}^{-1}(R_{1,2})=\Phi
_{2,2}^{-1}(I_{4})=\pm I_{4}$\smallskip 

\begin{tabular}{|l|l|l|l|}
	\hline
	$D=R_{3,4}$ & $\theta \in \left( 0,\pi \right) $ & $\theta \in \left( \pi
	,2\pi \right) $ & $\theta =\pi $ \\ \hline
	$\Phi _{2,2}^{-1}(D)$ & $\mp \left( 
	\begin{array}{cccc}
	\widehat{c} & 0 & 0 & \widehat{s} \\ 
	0 & \widehat{c} & -\widehat{s} & 0 \\ 
	0 & \widehat{s} & \widehat{c} & 0 \\ 
	-\widehat{s} & 0 & 0 & \widehat{c}%
	\end{array}%
	\right) $ & $\pm \left( 
	\begin{array}{cccc}
	\widehat{c} & 0 & 0 & \widehat{s} \\ 
	0 & \widehat{c} & -\widehat{s} & 0 \\ 
	0 & \widehat{s} & \widehat{c} & 0 \\ 
	-\widehat{s} & 0 & 0 & \widehat{c}%
	\end{array}%
	\right) $ & $\pm \left( 
	\begin{array}{cccc}
	0 & 0 & 0 & -1 \\ 
	0 & 0 & 1 & 0 \\ 
	0 & -1 & 0 & 0 \\ 
	1 & 0 & 0 & 0%
	\end{array}%
	\right) $ \\ \hline
\end{tabular}%
\smallskip 

For $\theta =0,2\pi ,$ $\Phi _{2,2}^{-1}(R_{3,4})=\Phi
_{2,2}^{-1}(I_{4})=\mp I_{4}$.\smallskip 

\begin{tabular}{|l|l|}
	\hline
	$D=H_{1,3}$ & $\Phi _{2,2}^{-1}(D)$ \\ \hline
	& $\mp \left( 
	\begin{array}{cccc}
	e^{\beta /2} & 0 & 0 & 0 \\ 
	0 & e^{-\beta /2} & 0 & 0 \\ 
	0 & 0 & e^{\beta /2} & 0 \\ 
	0 & 0 & 0 & e^{-\beta /2}%
	\end{array}%
	\right) $ \\ \hline
\end{tabular}%
\smallskip 

\begin{tabular}{|l|l|}
	\hline
	$D=H_{2,4}$ & $\Phi _{2,2}^{-1}(D)$ \\ \hline
	& $\mp \left( 
	\begin{array}{cccc}
	e^{\beta /2} & 0 & 0 & 0 \\ 
	0 & e^{\beta /2} & 0 & 0 \\ 
	0 & 0 & e^{-\beta /2} & 0 \\ 
	0 & 0 & 0 & e^{-\beta /2}%
	\end{array}%
	\right) $ \\ \hline
\end{tabular}
}
\end{theorem}

\noindent {\bf Proof:}
We just provide the details for the $H_{1,3}$ case. Following the
procedure outlined in Remark \ref{Outline}, we equate $\Phi_{2,2}(Y)$'s
entries to those of $H_{1,2}$. After a Groebner basis
calculation, the system of equations that have to be solved
are given by equating each member of $B$, as in the equation below,
to zero.
\[
B=\left\{ 
\begin{array}{c}
-1+\cosh (\beta )-\sinh (\beta ),y_{8}^{2}-(\cosh (\beta )-\sinh (\beta
)),y_{7},y_{6}-y_{8},y_{5}, \\ 
y_{4},y_{3}-(\cosh (\beta )+\sinh (\beta ))y_{8},y_{2},y_{1}-(\cosh (\beta
)+\sinh (\beta ))y_{8}
\end{array}
\right\}.
\]
Here $B$ is the set of equations provided by the procedure in Remark \ref
{Outline}. The ordering of variables for the lexicographic order is $%
y_{1}>y_{2}>\ldots >y_{8}$. Therefore, we have $y_{2}=y_{4}=y_{5}=y_{7}=0$, $y_{1}=\left( \cosh \beta +\sinh \beta \right) y_{8}$, $y_{3}=\left( \cosh
\beta +\sinh \beta \right) y_{8},$ $y_{6}=y_{8},$ $y_{8}^{2}=\left( \cosh
\left( \beta \right) -\sinh \left( \beta \right) \right) $. Solving this
system, we get
\[
y_{8}=\sqrt{\cosh (\beta ) -\sinh ( \beta ) }\ {\mbox or}\  
	y_{8}=-\sqrt{\cosh ( \beta ) -\sinh ( \beta ) }.
\]
Since
\[
\cosh ( \beta ) -\sinh ( \beta ) = e^{-\beta }\ 
	{\mbox and }\ \cosh ( \beta ) +\sinh ( \beta ) =e^{\beta }
\]
we have
\[
\begin{array}{ll}
y_{8}=\sqrt{\cosh ( \beta ) -\sinh ( \beta ) }=e
^{-\beta /2}, & y_{1}=( \cosh \beta +\sinh \beta ) y_{8}=e
^{\beta /2} \\ 
y_{3}= ( \cosh \beta +\sinh \beta ) y_{8}=e^{\beta /2}, & 
y_{6}=y_{8}=e^{-\beta /2}
\end{array}
\]
and analogously
\[
\begin{array}{ll}
y_{8}=-\sqrt{\cosh ( \beta ) -\sinh ( \beta ) }=-e
^{-\beta /2}, & y_{1}=( \cosh \beta +\sinh \beta ) y_{8}=-e
^{\beta /2} \\ 
y_{3}=( \cosh \beta +\sinh \beta ) y_{8}=-e^{\beta /2}, & 
y_{6}=y_{8}=-e^{-\beta /2}.
\end{array}
\]
Therefore, we have
\[
\Phi _{2,2}^{-1}(H_{1,3})=\mp \left( 
\begin{array}{cccc}
e^{\beta /2} & 0 & 0 & 0 \\ 
0 & e^{-\beta /2} & 0 & 0 \\ 
0 & 0 & e^{\beta /2} & 0 \\ 
0 & 0 & 0 & e^{-\beta /2}%
\end{array}
\right).\vspace{-0.1in} 
\] $\diamondsuit$

\section{Inversion of $\Phi_{3,2}$}\label{Section 5} 

The contents of Remark \ref{Outline} and Remark \ref{NotOnly} 
continue to be of pertinence to this section as well.

As before, by following the procedure in Example
\ref{Givens}, every matrix $X \in {\rm SO^{+}}(3, 2)$ can be 
represented non-uniquely
as a product of ordinary and hyperbolic Givens rotations as follows
\begin{equation}
\label{Givensfor32} 
D = R_{2,3}R_{1,2}R_{4,5}H_{1,4}R_{2,3}R_{4,5}H_{2,5}R_{4,5}H_{3,4}R_{4,5}. 
\end{equation} 

As in the previous section, we begin by detailing the
entries of $\Phi_{3,2} (Y)$, for any $Y\in {\mbox Spin^{+}}(3,2)$.

\noindent In accordance with \cite{fpi},
the basis of one-vectors for ${\mbox Cl}(3,2)$ is given by 
\begin{equation}
\label{Basis32}
{\mathbf B}_{3,2} = \{ X_{1}, X_{2}, X_{3}, X_{4},X_{5}\} 
\end{equation}
where

$%
\begin{array}{lll}
X_{1}=\sigma _{x}\otimes I_{4}, & X_{2}=\sigma _{z}\otimes \sigma 
_{x}\otimes I_{2}, & X_{3}=\sigma _{z}\otimes \sigma _{z}\otimes 
	\sigma _{z}, \\ 
X_{4}=(i \sigma _{y})\otimes I_{4}, & X_{5}=\sigma _{z}\otimes
(i \sigma _{y})\otimes I_{2}. & 
\end{array}%
$

\noindent Then, as shown in \cite{fpi}, ${\mbox Spin^{+}}(3,2)$ is the group 
\begin{equation}\label{32Spin}
\begin{array}{rcl}
{\mbox Spin}^{+}(3,2)&=& \{Y\in M(4, {\mathbf R});  Y^{T}MY = M\}
\end{array}
\end{equation} where
\[
M = \left (\begin{array}{cc}
0_{2} & {\rm J_{2}}\\
{\rm J_{2}} & 0_{2}
\end{array}
\right ), \enspace {\rm J_{2}} = \left (\begin{array}{cc}
0 & 1\\
-1 & 0
\end{array}
\right )
\]
viewed as living in $M(8, \mathbf{R})$ via the embedding
\begin{equation}\label{Embed32}
Y\rightarrow \hat{Y}
\end{equation}
where
\[
Y = \left (\begin{array}{cccc}
y_{1} & y_{2} & y_{3} & y_{4}\\
y_{5} & y_{6} & y_{7} & y_{8}\\
y_{9} & y_{10} & y_{11} & y_{12}\\
y_{13} & y_{14} & y_{15} & y_{16}
\end{array}
\right )
\]
and
\[
\hat{Y}=\left( 
\begin{array}{cccccccc}
y_{1} & 0 & y_{2} & 0 & y_{3} & 0 & y_{4} & 0 \\ 
0 & y_{1} & 0 & -y_{2} & 0 & -y_{3} & 0 & y_{4} \\ 
y_{5} & 0 & y_{6} & 0 & y_{7} & 0 & y_{8} & 0 \\ 
0 & -y_{5} & 0 & y_{6} & 0 & y_{7} & 0 & -y_{8} \\ 
y_{9} & 0 & y_{10} & 0 & y_{11} & 0 & y_{12} & 0 \\ 
0 & -y_{9} & 0 & y_{10} & 0 & y_{11} & 0 & -y_{12} \\ 
y_{13} & 0 & y_{14} & 0 & y_{15} & 0 & y_{16} & 0 \\ 
0 & y_{13} & 0 & -y_{14} & 0 & -y_{15} & 0 & y_{16}%
\end{array}
\right).
\]

\begin{remark}
{\rm The group defined by Equation (\ref{32Spin}) is explicitly
conjugate to the standard representation of the real symplectic
group $Sp(4, \mathbf{R})$. 
This conjugation can be found 
without any eigencalculations. 
By this we mean that the $4\times 4$ skew-symmetric matrix  $M$ 
of Equation (\ref{32Spin})
defining this version of $Sp (4, \mathbf{R})$ can be rendered explicitly, 
conjugate to $J_{4}$, without any eigencalculations. We omit the
details.}
\end{remark}
 
\noindent  Next, entries of $Y$ (and thus, also those
of $\hat{Y}$) satisfy quadratic relations emanating from
Equation (\ref{32Spin}) which read as follows

\qquad $f_{1}{=y}_{{1}}{y}_{{16}}{-y}_{4}y_{13}{-y}_{5}{y}_{12}{+y}_{8}{y}%
_{9}{-1=0}$

\qquad $f_{2}{=y}_{{2}}{y}_{{16}}{-y}_{{4}}{y}_{{14}}{-y}_{{6}}{y}_{{12}}{%
+y}_{{8}}{y}_{{10}}{=0}$

\qquad $f_{3}{=y}_{{3}}{y}_{{16}}{-y}_{{4}}{y}_{{15}}{-y}_{{7}}{y}_{{12}}{%
+y}_{{8}}{y}_{{11}}={0}$

\qquad $f_{4}{=y}_{{1}}{y}_{{15}}{-y}_{{3}}{y}_{{13}}{-y}_{{5}}{y}_{{11}}{%
+y}_{{7}}y_{{9}}={0}$

\qquad $f_{5}{=y}_{{2}}{y}_{{15}}{-y}_{{3}}{y}_{{14}}{-y}_{{6}}{y}_{{11}}{%
+y}_{{7}}{y}_{{10}}{+1=0}$

\qquad $f_{6}{=y}_{{1}}{y}_{{14}}{-y}_{{2}}{y}_{{13}}{-y}_{{5}}{y}_{{10}}{%
+y}_{{6}}{y}_{{9}}=0$.\vspace*{0.10in}

As outlined in Remark \ref{NotOnly},
we use a Groebner basis for the ideal spanned by the polynomials
$\{f_{1}, \ldots , f_{6}\}$ with respect to the lexicographic order,
to calculate the matrix of the linear map $V\rightarrow YVY^{-1}$,
where $V$ is a one-vector and $Y\in {\mbox Spin^{+}}(3,2)$,
with respect to the basis ${\mathbf B}=
\{ X_{1}, X_{2}, X_{3}, X_{4}, X_{5}\}$, from Equation (\ref{Basis32}),
Then this matrix equals $M=(m_{i,j}), i,j=1,2,3,4,5$. The $m_{i,j}$ are given by
\begin{small}
\begin{eqnarray}
&&
\begin{tabular}{lll}
$m_{1,1}$ & $=$ & $\frac{1}{2}%
(-y_{10}y_{13}-y_{12}y_{15}+y_{11}y_{16}-y_{2}y_{5}+y_{1}y_{6}-
y_{4}y_{7}+y_{3}y_{8}+y_{14}y_{9}) 
$ \\ 
$m_{2,1}$ & $=$ & $\frac{1}{2}%
(-y_{1}y_{10}-y_{12}y_{3}+y_{11}y_{4}+y_{14}y_{5}-
y_{13}y_{6}+y_{16}y_{7}-y_{15}y_{8}+y_{2}y_{9}) 
$ \\ 
$m_{3,1}$ & $=$ & $y_{10}y_{5}+y_{12}y_{7}-y_{11}y_{8}-y_{6}y_{9}$ \\ 
$m_{4,1}$ & $=$ & $\frac{1}{2}%
(y_{10}y_{13}+y_{12}y_{15}-y_{11}y_{16}-y_{2}y_{5}+
y_{1}y_{6}-y_{4}y_{7}+y_{3}y_{8}-y_{14}y_{9}) 
$ \\ 
$m_{5,1}$ & $=$ & $\frac{1}{2}%
(-y_{1}y_{10}-y_{12}y_{3}+y_{11}y_{4}-y_{14}y_{5}+y_{13}y_{6}-y_{16}y_{7}+
y_{15}y_{8}+y_{2}y_{9}) 
$%
\end{tabular}
\\
&&%
\begin{tabular}{lll}
$m_{1,2}$ & $=$ & $\frac{1}{2}%
(y_{11}y_{13}-y_{12}y_{14}+y_{10}y_{16}+y_{3}y_{5}-y_{4}y_{6}-y_{1}y_{7}+
y_{2}y_{8}-y_{15}y_{9}) 
$ \\ 
$m_{2,2}$ & $=$ & $\frac{1}{2}%
(y_{1}y_{11}-y_{12}y_{2}+y_{10}y_{4}-y_{15}y_{5}+y_{16}y_{6}+y_{13}y_{7}-
y_{14}y_{8}-y_{3}y_{9}) 
$ \\ 
$m_{3,2}$ & $=$ & $-y_{11}y_{5}+y_{12}y_{6}-y_{10}y_{8}+y_{7}y_{9}$ \\ 
$m_{4,2}$ & $=$ & $\frac{1}{2}%
(-y_{11}y_{13}+y_{12}y_{14}-y_{10}y_{16}+y_{3}y_{5}-y_{4}y_{6}-
y_{1}y_{7}+y_{2}y_{8}+y_{15}y_{9}) 
$ \\ 
$m_{5,2}$ & $=$ & $\frac{1}{2}%
(y_{1}y_{11}-y_{12}y_{2}+y_{10}y_{4}+y_{15}y_{5}-y_{16}y_{6}-y_{13}y_{7}+
y_{14}y_{8}-y_{3}y_{9}) 
$%
\end{tabular}
\end{eqnarray}
\end{small}
\begin{small}
\begin{eqnarray}
&&
\begin{tabular}{lll}
$m_{1,3}$ & $=$ & $-y_{11}y_{14}+y_{10}y_{15}-y_{3}y_{6}+y_{2}y_{7}$ \\ 
$m_{2,3}$ & $=$ & $-y_{11}y_{2}+y_{10}y_{3}+y_{15}y_{6}-y_{14}y_{7}$ \\ 
$m_{3,3}$ & $=$ & $-1+2y_{11}y_{6}-2y_{10}y_{7}$ \\ 
$m_{4,3}$ & $=$ & $y_{11}y_{14}-y_{10}y_{15}-y_{3}y_{6}+y_{2}y_{7}$ \\ 
$m_{5,3}$ & $=$ & $-y_{11}y_{2}+y_{10}y_{3}-y_{15}y_{6}+y_{14}y_{7}$%
\end{tabular}\ \\
&&
\begin{tabular}{lll}
$m_{1,4}$ & $=$ & $\frac{1}{2}%
(-y_{10}y_{13}+y_{12}y_{15}-y_{11}y_{16}-y_{2}y_{5}+y_{1}y_{6}+y_{4}y_{7}-
y_{3}y_{8}+y_{14}y_{9}) 
$ \\ 
$m_{2,4}$ & $=$ & $\frac{1}{2}%
(-y_{1}y_{10}+y_{12}y_{3}-y_{11}y_{4}+y_{14}y_{5}-y_{13}y_{6}-y_{16}y_{7}+
y_{15}y_{8}+y_{2}y_{9}) 
$ \\ 
$m_{3,4}$ & $=$ & $y_{10}y_{5}-y_{12}y_{7}+y_{11}y_{8}-y_{6}y_{9}$ \\ 
$m_{4,4}$ & $=$ & $\frac{1}{2}%
(y_{10}y_{13}-y_{12}y_{15}+y_{11}y_{16}-y_{2}y_{5}+y_{1}y_{6}+y_{4}y_{7}-
y_{3}y_{8}-y_{14}y_{9}) 
$ \\ 
$m_{5,4}$ & $=$ & $\frac{1}{2}%
(-y_{1}y_{10}+y_{12}y_{3}-y_{11}y_{4}-y_{14}y_{5}+y_{13}y_{6}+y_{16}y_{7}-
y_{15}y_{8}+y_{2}y_{9}) 
$%
\end{tabular}
\\
&&%
\begin{tabular}{lll}
$m_{1,5}$ & $=$ & $\frac{1}{2}%
(y_{11}y_{13}+y_{12}y_{14}-y_{10}y_{16}+y_{3}y_{5}+y_{4}y_{6}-
y_{1}y_{7}-y_{2}y_{8}-y_{15}y_{9}) 
$ \\ 
$m_{2,5}$ & $=$ & $\frac{1}{2}%
(y_{1}y_{11}+y_{12}y_{2}-y_{10}y_{4}-y_{15}y_{5}-y_{16}y_{6}+y_{13}y_{7}+
y_{14}y_{8}-y_{3}y_{9}) 
$ \\ 
$m_{3,5}$ & $=$ & $-y_{11}y_{5}-y_{12}y_{6}+y_{10}y_{8}+y_{7}y_{9}$ \\ 
$m_{4,5}$ & $=$ & $\frac{1}{2}%
(-y_{11}y_{13}-y_{12}y_{14}+y_{10}y_{16}+y_{3}y_{5}+y_{4}y_{6}-y_{1}y_{7}-
y_{2}y_{8}+y_{15}y_{9}) 
$ \\ 
$m_{5,5}$ & $=$ & $\frac{1}{2}
(y_{1}y_{11}+y_{12}y_{2}-y_{10}y_{4}+y_{15}y_{5}+y_{16}y_{6}-y_{13}y_{7}-
y_{14}y_{8}-y_{3}y_{9}.$
\end{tabular}
\end{eqnarray}
\end{small}

Thus, finding the preimage of $X\in SO^{+}(3,2,\mathbf{R})$, means
solving for the 16 unknowns $y_{1}, \ldots , y_{16}$ from the 25 equations
$m_{ij} = x_{ij}$, where $x_{ij}$ are the entries of $X$.

We now let $X$ be one of the six \underline{distinct}
$R_{ij}(c, s)$ or $H_{kl}(a, b)$ where the
$R_{ij}$ and $H_{kl}$ are as in Equation (\ref{Givensfor32}).
The next result gives explicit formulae
for $y_{1}$, $y_{2}$, ..., $y_{16}$ in terms of $s$, $c$, $a$, $b$.

\begin{theorem}\label{32Solutions}
{\rm Let $c = \cos \theta $, 
$s = \sin \theta$, $a = \cosh\beta $, $b= \sinh \beta$.
Correspondingly, let $\widehat{c} = \cos \frac{\theta}{2}$,
$\widehat{s} = \sin \frac{\theta}{2}$, $\widehat{ch}
= \cosh \frac{\beta}{2}$, $\widehat{sh} =
\sinh  \frac{\beta}{2}$.
Then, the preimages $Y\in {\mbox Spin^{+}}(3,2)$ of the Givens  
factors in Equation (\ref{Givensfor32}) are\ \\
\begin{tabular}{|l|l|l|l|}
	\hline
	$M=R_{1,2}$ & $\theta \in \left( 0,\pi \right) $ & $\theta \in \left( \pi
	,2\pi \right) $ & $\theta =\pi $ \\ \hline
	$\Phi _{3,2}^{-1}(M)$ & $\mp \left( 
	\begin{array}{cccc}
	\widehat{c} & 0 & 0 & \widehat{s} \\ 
	0 & \widehat{c} & \widehat{s} & 0 \\ 
	0 & -\widehat{s} & \widehat{c} & 0 \\ 
	-\widehat{s} & 0 & 0 & \widehat{c}%
	\end{array}%
	\right) $ & $\pm \left( 
	\begin{array}{cccc}
	\widehat{c} & 0 & 0 & \widehat{s} \\ 
	0 & \widehat{c} & \widehat{s} & 0 \\ 
	0 & -\widehat{s} & \widehat{c} & 0 \\ 
	-\widehat{s} & 0 & 0 & \widehat{c}%
	\end{array}%
	\right) $ & $\pm \left( 
	\begin{array}{cccc}
	0 & 0 & 0 & -1 \\ 
	0 & 0 & -1 & 0 \\ 
	0 & 1 & 0 & 0 \\ 
	1 & 0 & 0 & 0%
	\end{array}%
	\right) $ \\ \hline
\end{tabular}%
\smallskip 

For $\theta =0,2\pi$\quad $\Phi _{3,2}^{-1}(R_{1,2})=\Phi
_{3,2}^{-1}(I_{4})=\pm I_{4}.$\smallskip 

\begin{tabular}{|l|l|l|l|}
	\hline
	$M=R_{1,3}$ & $\theta \in \left( 0,\pi \right) $ & $\theta \in \left( \pi
	,2\pi \right) $ & $\theta =\pi $ \\ \hline
	$\Phi _{3,2}^{-1}(M)$ & $\mp \left( 
	\begin{array}{cccc}
	\widehat{c} & 0 & \widehat{s} & 0 \\ 
	0 & \widehat{c} & 0 & -\widehat{c} \\ 
	-\widehat{s} & 0 & \widehat{c} & 0 \\ 
	0 & \widehat{s} & 0 & \widehat{c}%
	\end{array}%
	\right) $ & $\pm \left( 
	\begin{array}{cccc}
	\widehat{c} & 0 & \widehat{s} & 0 \\ 
	0 & \widehat{c} & 0 & -\widehat{c} \\ 
	-\widehat{s} & 0 & \widehat{c} & 0 \\ 
	0 & \widehat{s} & 0 & \widehat{c}%
	\end{array}%
	\right) $ & $\pm \left( 
	\begin{array}{cccc}
	0 & 0 & 1 & 0 \\ 
	0 & 0 & 0 & -1 \\ 
	-1 & 0 & 0 & 0 \\ 
	0 & 1 & 0 & 0%
	\end{array}%
	\right) \hspace{-0.01in}$ \\ \hline
\end{tabular}%
\smallskip 

For $\theta =0,2\pi$, \quad $\Phi _{3,2}^{-1}(R_{1,3})=\Phi
_{3,2}^{-1}(I_{4})=\pm I_{4}.$\smallskip \\
\begin{small}
\begin{tabular}{|l|l|l|l|}
	\hline
	$M=R_{2,3}$ & $\theta \in \left( 0,\pi \right) $ & $\theta \in \left( \pi
	,2\pi \right) $ & $\theta =\pi $ \\ \hline
	$\Phi _{3,2}^{-1}(M)$ & $\mp \left( 
	\begin{array}{cccc}
	\widehat{c} & \widehat{s} & 0 & 0 \\ 
	-\widehat{s} & \widehat{c} & 0 & 0 \\ 
	0 & 0 & \widehat{c} & -\widehat{s} \\ 
	0 & 0 & -\widehat{s} & \widehat{c}%
	\end{array}%
	\right) $ & $\pm \left( 
	\begin{array}{cccc}
	\widehat{c} & \widehat{s} & 0 & 0 \\ 
	-\widehat{s} & \widehat{c} & 0 & 0 \\ 
	0 & 0 & \widehat{c} & -\widehat{s} \\ 
	0 & 0 & -\widehat{s} & \widehat{c}%
	\end{array}%
	\right) $ & $\pm \left( 
	\begin{array}{cccc}
	0 & 1 & 0 & 0 \\ 
	-1 & 0 & 0 & 0 \\ 
	0 & 0 & 0 & 1 \\ 
	0 & 0 & -1 & 0%
	\end{array}%
	\right) \hspace{0.10in}$ \\ \hline
\end{tabular}%
\smallskip 

For $\theta =0,2\pi$, \quad  $\Phi _{3,2}^{-1}(R_{2,3})=\Phi
_{3,2}^{-1}(I_{4})=\pm I_{4}.$\smallskip 

\begin{tabular}{|l|l|l|l|}
	\hline
	$M=R_{4,5}$ & $\theta \in \left( 0,\pi \right) $ & $\theta \in \left( \pi
	,2\pi \right) $ & $\theta =\pi $ \\ \hline
	$\Phi _{3,2}^{-1}(M)$ & $\mp \left( 
	\begin{array}{cccc}
	\widehat{c} & 0 & 0 & -\widehat{s} \\ 
	0 & \widehat{c} & \widehat{s} & 0 \\ 
	0-\widehat{s} & \widehat{c} & 0 &  \\ 
	\widehat{s} & 0 & 0 & \widehat{c}%
	\end{array}%
	\right) $ & $\pm \left( 
	\begin{array}{cccc}
	\widehat{c} & 0 & 0 & -\widehat{s} \\ 
	0 & \widehat{c} & \widehat{s} & 0 \\ 
	0 & -\widehat{s} & \widehat{c} & 0 \\ 
	\widehat{s} & 0 & 0 & \widehat{c}%
	\end{array}%
	\right) $ & $\pm \left( 
	\begin{array}{cccc}
	0 & 0 & 0 & -1 \\ 
	0 & 0 & 1 & 0 \\ 
	0 & -1 & 0 & 0 \\ 
	1 & 0 & 0 & 0%
	\end{array}%
	\right) \hspace{0.12in}$ \\ \hline
\end{tabular}%
\smallskip 

For $\theta =0,2\pi$,\quad $\Phi _{3,2}^{-1}(R_{4,5})=\Phi
_{3,2}^{-1}(I_{4})=\pm I_{4}$.\smallskip 
\end{small}\\
\begin{tabular}{|l|l|}
	\hline
	$M=H_{1,4}$ & $\Phi _{3,2}^{-1}(M)$ \\ \hline
	& $\mp \left( 
	\begin{array}{cccc}
	e^{\beta /2} & 0 & 0 & 0 \\ 
	0 & e^{\beta /2} & 0 & 0 \\ 
	0 & 0 & e^{-\beta /2} & 0 \\ 
	0 & 0 & 0 & e^{-\beta /2}%
	\end{array}%
	\right) $ \\ \hline
\end{tabular}%
\smallskip 

\begin{tabular}{|l|l|}
	\hline
	$M=H_{2,5}$ & $\Phi _{2,2}^{-1}(M)$ \\ \hline
	& $\mp \left( 
	\begin{array}{cccc}
	e^{\beta /2} & 0 & 0 & 0 \\ 
	0 & e^{-\beta /2} & 0 & 0 \\ 
	0 & 0 & e^{\beta /2} & 0 \\ 
	0 & 0 & 0 & e^{-\beta /2}%
	\end{array}%
	\right) $ \\ \hline
\end{tabular}%
\smallskip 

\begin{tabular}{|l|l|}
	\hline
	$M=H_{3,4}$ & $\Phi _{2,2}^{-1}(M)$ \\ \hline
	& $\mp \left( 
	\begin{array}{cccc}
	\widehat{ch} & 0 & -\widehat{sh} & 0 \\ 
	0 & \widehat{ch} & 0 & \widehat{sh} \\ 
	-\widehat{sh} & 0 & \widehat{ch} & 0 \\ 
	0 & \widehat{sh} & 0 & \widehat{ch}%
	\end{array}%
	\right) \hspace{0.451in}$ \\ \hline
\end{tabular}
}
\end{theorem}

\noindent {\bf Proof:} We provide the details for $H_{1,4}$.
The procedure in Remark \ref{Outline} yields the following
system of equations
\begin{eqnarray*}
	B &=&\{-1+\cosh ^{2}(\beta )-\sinh ^{2}(\beta ),x_{16}^{2}-(\cosh (\beta
	)-\sinh (\beta )),x_{15},x_{14},x_{13}, \\
	&&\enspace x_{12},x_{11}-x_{16},x_{10},x_{9},x_{8},x_{7},x_{6}-(\cosh (\beta
	)+\sinh (\beta ))x_{16},x_{5},x_{4},x_{3}, \\
	&&\enspace x_{2},x_{1}-(\cosh (\beta )+\sinh (\beta ))x_{16}\}.
\end{eqnarray*}
Therefore, we have
\[
x_{2}=x_{3}=x_{4}=x_{5}=x_{7}=x_{8}=x_{9}=x_{10}=x_{15}=x_{14}=x_{13}=x_{12}=0
\]
and
\[
\begin{tabular}{ll}
$x_{11}=x_{16},$ & $x_{6}=\left( \cosh (\beta )+\sinh (\beta )\right) x_{16},
$ \\ 
$x_{1}=\left( \cosh (\beta )+\sinh (\beta )\right) x_{16},$ & $
x_{16}^{2}=\cosh (\beta )-\sinh (\beta ).$
\end{tabular}
\]
Solving this system, we get
\[
x_{16}=\sqrt{\cosh (\beta )-\sinh (\beta )}\  {\mbox or }\ 
 x_{16}=-\sqrt{\cosh(\beta )-\sinh (\beta )}.
\]
Since
\[
\cosh (\beta )-\sinh (\beta )=e^{-\beta }\ {\mbox and }\  
\cosh (\beta )+\sinh(\beta )=e^{\beta }
\]
and if $x_{16}=\sqrt{\cosh ( \beta ) -\sinh ( \beta ) }
=e^{-\beta /2},$ we get $x_{1}=e^{\beta /2},$ $x_{6}=e^{\beta /2}$ and $%
x_{11}=x_{16}=e^{-\beta /2}$. Analogously, if 
$x_{16}=-\sqrt{\cosh (\beta )
	-\sinh (\beta )}-e^{-\beta /2},$ 
then $x_{1}=-e^{\beta /2},$ $x_{6}=-e%
^{\beta /2}$ and $x_{11}=x_{16}=-e^{-\beta /2}$. Therefore, we have

\[
\Phi _{3,2}^{-1}(H_{1,4})=\mp \left( 
\begin{array}{cccc}
e^{\beta /2} & 0 & 0 & 0 \\ 
0 & e^{\beta /2} & 0 & 0 \\ 
0 & 0 & e^{-\beta /2} & 0 \\ 
0 & 0 & 0 & e^{-\beta /2}%
\end{array}%
\right). 
\] $\diamondsuit$ 

\section{Inversion of $\Phi_{4,1}$ via the Inversion of $\Psi_{4,1}$}\label{Section 6}
In this section the map $\Phi_{4,1}: {\mbox Spin}^{+}(4,1) \rightarrow
{\rm SO^{+}}(4,1)$ is inverted by linearizing $\Phi_{4,1}$. We will see
that this modus operandi works elegantly for both the case where
the target matrix in ${\rm SO^{+}}(4,1)$ is assumed to be given by its Givens
factors and the case wherein we assume that the target matrix is
given by its polar decomposition. In particular, we will see
that the latter provides a constructive technique to find the
polar decomposition of a matrix in ${\mbox Spin^{+}}(4,1)$. Since this
a group of certain $2\times 2$ quaternionic matrices, we have thus
a technique to compute the polar decomposition of such quaternionic
matrices, without passage to the associated $\Theta_{{\mathbf H}}$ image in
$M(4, {\mathbf C})$ and, in particular, without any eigencalculations.

As usual we begin with a basis of one-vectors for ${\mbox Cl}(4,1) =
M(4, \mathbf{C})$
\[
V_{1} = \sigma_{z}\otimes \sigma_{x}, V_{2} = \sigma_{y}\otimes I_{2},
	V_{3} = \sigma_{z}\otimes \sigma_{z}, V_{4} = \sigma_{x}\otimes I_{2},
V_{5} = -\sigma_{z}\otimes (i\sigma_{y}).
\]
As shown in \cite{fpi}, with respect to this basis,
$$
{\mbox Spin}^{+}(4,1) = \{X \in M(4, C)\cap {\mbox Im} 
(\Theta_{H});  X^{*}MX = M\}
$$
where
\[
 M = (i\sigma_{y}) \oplus (-i\sigma_{y}).
\]

Since these matrices are $\Theta_{H}$ matrices it is convenient
to identify them with the corresponding matrices in $M(2, {\mathbf H})$.
Note, however, $M$ itself is not a $\Theta_{{\mathbf H}}$ matrix. 

Next, the Lie algebra of the spin group equals 
$$
{\mbox spin}^{+}(4, 1) = \{
\Lambda \in M(4, C)\cap {\mbox Im}(\Theta_{{\mathbf H}});
\Lambda^{*}M = - M\Lambda \}
$$

Since $\Lambda$ is in the image of $\Theta_{H}$ it
is of the form $\left (\begin{array}{cc} Z & W\\
-\bar{W} & \bar{Z}
\end{array}
\right )$ with $Z + W j\in M(2, \mathbf{H})$. The condition
$\Lambda^{*}M =  -M\Lambda$ forces
\begin{equation}
\label{EquationZ}
Z = \left (\begin{array}{cc}
a_{1} + i a_{2} & b\\
c & -a_{1} + i a_{2}
\end{array}
\right )
\end{equation} 
and
\begin{equation}
\label{EquationW} 
W = \left (\begin{array}{cc}
\alpha_{1} + i\alpha_{2} & \beta_{1} + i\beta_{2}\\
\gamma_{1} + i\gamma_{2} & -\alpha_{1} - i\alpha_{2}
\end{array}
\right ).
\end{equation} 

So
\[
\Lambda =
\left(
\begin{array}{cccc}
 a_1+i a_2 & b & \alpha _1+ i \alpha _2 & \beta _1+i \beta _2 \\
 c & -a_1+i a_2 & \gamma _1+i \gamma _2 & -\alpha _1-i \alpha _2 \\
 -\alpha _1+i \alpha _2 & -\beta _1+i \beta _2 & a_1-i a_2 & b \\
 -\gamma _1+i \gamma _2 & \alpha _1-i \alpha _2 & c & -a_1-i a_2 \\
\end{array}
\right).
\]

The linearization of $\Phi_{4,1}$ associates to
$\Lambda\in {\mbox spin}^{+}(4,1)$
the matrix of the linear map which sends a one-vector $V$ to the
one-vector $YV - VY$, with respect to the basis
$\{V_{1}, \ldots , V_{5}\}$ above. We then get
\begin{equation}\label{ImPsi41}
\Psi_{4,1}(\Lambda )
= \left(
\begin{array}{ccccc}
 0 &  \beta _2+ \gamma _2 & - b+ c &  \beta _1+ \gamma _1 & -2 a_1 \\
 - \beta _2- \gamma _2 & 0 & -2 \alpha _2 & 2 a_2 & - \beta _2+ \gamma _2 \\
  b- c & 2 \alpha _2 & 0 & 2 \alpha _1 &  b+ c \\
 - \beta _1- \gamma _1 & -2 a_2 & -2 \alpha _1 & 0 & - \beta _1+ \gamma _1 \\
 -2 a_1 & - \beta _2+ \gamma _2 &  b+ c & - \beta _1+ \gamma _1 & 0 
\end{array}
\right).
 \end{equation}

\subsection{Inversion via Givens Factors}\label{Section 6.1}
Following Example \ref{Givens} every 
matrix in $SO^{+}(4,1)$ can be decomposed non-uniquely as
\[
X = R_{14}R_{13}R_{12}H_{15}R_{24}R_{23}H_{25}R_{34}H_{35}H_{45}. 
\]

We then have the following result.

\begin{proposition}\label{Phi41InversionViaGivens}
{\rm The table immediately below
describes the $Y \in {\mbox Spin}^{+}(4,1)
\subseteq M(2, \mathbf{H})$ satisfying $\Phi_{4,1}(\pm Y )
= X$, where $X$ is one of the Givens matrices in the last equation.}
\end{proposition}

\noindent {\bf Proof:} The proof proceeds by expressing
each $R_{ij}$ or $H_{ij}$ as the exponential of an
$L_{ij}\in so(5,1)$, finding the $2\times 2$ quaternionic matrix
$K_{ij} = \Psi^{-1}_{4,1} (L_{ij})$ and then exponentiating
$K_{ij}$ explicitly. This last matrix is $Y$.
For brevity only the details for $H_{25}$ are displayed.

We begin by noting that
$H_{25} = {\mbox Exp} [\theta (e_{2}e_{5}^{T} + e_{5}e_{2}^{T})]$.
By inspecting, Equation (\ref{ImPsi41}),
it is seen that its preimage in ${\mbox spin}^{+}(4,1)$ is
\[
\Theta_{{\mathbf H}} [\frac{\theta}{2}\left (\begin{array}{cc}
0 &  -k\\
k & 0
\end{array}
\right ) ].
\]

Now
\[
[\frac{\theta}{2}\left (\begin{array}{cc}
0 &  -k\\
k &   0
\end{array}
\right )]^{2} = 
\frac{\theta^{2}}{4}I_{2}.
\]
Hence \[
{\mbox Exp} [\frac{\theta}{2}\left (\begin{array}{cc}
0 &  -k\\
k & 0
\end{array}
\right )]  
 =\left (\begin{array}{cc}
\cosh (\theta /2) & -\sinh (\theta /2)k\\
\sinh (\theta /2)k & \cosh (\theta /2)
\end{array}
\right )\diamondsuit .
\]

\begin{center}
\begin{tabular}{|c|c|} 
\hline
$R_{ij}$ {or} $H_{ij}$ & $Y $ \\ \hline
$R_{14}$ & $\left( 
\begin{array}{cc}
\cos (\theta /2) & -\sin (\theta /2)j \\ 
-\sin (\theta /2)j & \cos (\theta /2)%
\end{array}%
\right) $ \\ \hline
$R_{13}$ & $\left( 
\begin{array}{cc}
\cos (\theta /2) & -\sin (\theta /2) \\ 
\sin (\theta /2) & \cos (\theta /2)%
\end{array}%
\right) $ \\ \hline
$R_{1,2}$ & $\left( 
\begin{array}{cc}
\cos (\theta /2) & -\sin (\theta /2)k \\ 
-\sin (\theta /2)k & \cos (\theta /2)%
\end{array}%
\right) $ \\ \hline
$H_{1,5}$ & $\left( 
\begin{array}{cc}
e^{-\theta /2} & 0 \\ 
0 & e^{-\theta /2}%
\end{array}%
\right) $ \\ \hline
$R_{4,2}$ & $\left( 
\begin{array}{cc}
\cos (\theta /2)+\sin (\theta /2)i  & 0 \\ 
0 & \cos (\theta /2)-\sin (\theta /2)i 
\end{array}%
\right) $ \\ \hline
$R_{3,2}$ & $\left( 
\begin{array}{cc}
\cos (\theta /2)-\sin (\theta /2)k & 0 \\ 
0 & \cos (\theta /2)+\sin (\theta /2)k%
\end{array}%
\right) $ \\ \hline
$H_{2,5}$ & $\left( 
\begin{array}{cc}
\cosh (\theta /2) & -\sinh (\theta /2)k \\ 
\sinh (\theta /2)k & \cosh (\theta /2)%
\end{array}%
\right) $ \\ \hline
$H_{3,5}$ & $\left( 
\begin{array}{cc}
\cosh (\theta /2) & \sinh (\theta /2)k \\ 
\sinh (\theta /2)k & \cosh (\theta /2)%
\end{array}%
\right) $ \\ \hline
$H_{4,5}$ & $\left( 
\begin{array}{cc}
\cosh (\theta /2) & -\sinh (\theta /2)j \\ 
\sinh (\theta /2)j & \cosh (\theta /2)%
\end{array}%
\right) $ \\ \hline
\end{tabular}
\end{center}

\begin{remark}\label{agnostic}
Agnostic Inversion - Linearization and Givens in General: 
{\rm The fact that the preimages of the logarithms of 
the Givens factors in the spin 
Lie algebra always had a quadratic 
minimal polynomial holds for general $(p, q)$ case. 
This provides us with a method to 
invert both the abstract $\Phi_{p, q}$ and the matrix 
$\Phi_{p, q}$ without having to find a concrete form of $\Phi_{p,q}$.
We dub the latter as \underline{agnostic inversion}. 
Thus, the method of \cite{shiro}
as enhanced by iii) of Remark \ref{nuis} is agnostic inversion. 
We will now justify our claim
and thereby display a second method 
for agnostic inversion which uses Givens decompositions instead
of calculating minors of $X$.

Specifically, the spin Lie algebra is also the space of bivectors. Let $L_{ij}
= \theta (e_{i}e_{j}^{T} + e_{j}e_{i}^{T})$ be the logarithm of a hyperbolic Givens
$H_{ij}$. Its preimage in the space of bivectors is $\frac{\theta}{2}X_{i}X_{j}$, where $1\leq p \leq p$ and $p+1\leq j \leq q$. Indeed the abstract $\Psi_{p, q}$ sends an element $\Lambda$ in
the space of bivectors to the matrix of the linear map which sends a one-vector $V$ to
$\Lambda V - V\Lambda$. From the form of $L_{ij}$ it then follows that if $\Lambda$ is the preimage of $L_{ij}$, then $\Lambda$ commutes with all one-vectors in the basis of
one-vectors $\{X_{1}, \ldots, X_{p}, X_{p+1}, \ldots, X_{q}\}$ except $X_{i}$ and $X_{j}$. This
observation plus a few calculations show that $\Lambda =  \frac{\theta}{2}X_{i}X_{j}$. Quite clearly,
$\Lambda^{2}$ is a positive multiple of the identity of ${\mbox Cl}(p, q)$. 
So 
${\mbox Exp}(\Lambda)=\cosh(\frac{\theta}{2})I +\sinh(\frac{\theta}{2})
[X_{i}X_{j}]$ is the preimage of $H_{ij}$ in the spin Lie group. Similar comments apply to $R_{ij}$. This provides the inversion of the abstract covering map
and also the agnostic inversion of $\Phi_{p, q}$ 
via iii) of Remark \ref{nuis}. 

For the inversion  
of the \underline {concrete} $\Phi_{p, q}$ via linearization 
we need, of course, an explicit matrix form of $\Psi_{p, q}$. 
However, since the embedding of the even sublagebra 
in the full ${\mbox Cl}(p,q)$ is an algebra isomorphism onto its image, it 
is guaranteed that the $\Psi_{p,q}^{-1}(L_{ij})$ also satisfies 
the same quadratic annihilating polynomial and hence 
its exponential is easily found.}
\end{remark}

\subsection{Inversion of $\Phi_{4,1}$ via the polar decomposition}\label{Section 6.2}

Let $X\in {\rm SO^{+}}(4, 1)$. Then in view of Remark \ref{PDforn1},
one can find constructively both its polar decomposition \[
X = VP\]
and the $\hat{X}\in so(4,1)$ , such that it is symmetric and 
${\mbox Exp} (\hat{X}) = P$. Furthermore, by invoking Remark \ref{FindingLog} plus
a little work, we can also find a skew-symmetric, $5\times 5$, real matrix
whose exponential equals $V$.
 
We will presently see that it is possible to exponentiate in closed form
the preimage, under $\Psi_{4,1}$, of a symmetric matrix or a skew-symmetric
matrix in
$so(4,1)$. Therefore, using the polar decomposition to
invert $\Phi_{4,1}$ is a viable option.

To that end let $\hat{X}\in so(4,1)$ be symmetric. Then its preimage in
${\mbox spin}^{+}(4,1)$ is the $\Theta_{{\mathbf H}}$ image of the following
$2\times 2$ quaternionic matrix
\begin{equation}
\label{LambdaSymmetric}
\Lambda = \left (\begin{array}{cc}
a_{1} & b + \beta_{1}j + \beta_{2}k\\
b - \beta_{1}j - \beta_{2}k & -a_{1}
\end{array}
\right ).
\end{equation}
 
A quick calculation $\Lambda^{2} = \lambda^{2} I_{2}$ where

\begin{equation}
\label{lambda}
\lambda^{2} = a_{1}^{2} + \mid q\mid^{2}
\end{equation}
 wherein $q$ is the quaternion
$b + \beta_{1}j + \beta_{2}k$. Therefore
$${\mbox Exp} (Y) = \cosh (\lambda )I_{2} 
+ \frac{\sinh (\lambda )}{\lambda}\Lambda.$$
Hence, the preimage of $P = {\mbox Exp}(\hat{X})$ is
$\Theta_{{\mathbf H}} 
[  \cosh (\lambda )I_{2} + \frac{\sinh (\lambda )}{\lambda}
\Lambda ]$.  
Note also that
\[
[  \cosh (\lambda )I_{2} + \sinh (\lambda )\Lambda ]^{-1}
= \cosh (\lambda )I_{2} - \sinh (\lambda )\Lambda .
\]

Next, $V$ is both special orthogonal and in ${\rm SO^{+}}(4, 1)$. Therefore,
it is of the form $W\oplus 1$, where $W$ is $4\times 4$ special orthogonal.
Hence the matrix $\hat{Y} \in so(4, 1)$ with 
\[
{\mbox Exp}(\hat{Y}) = V 
\]

is of the form
\[
\hat{Y} = Y\oplus 0_{1\times 1}
\]
with $Y$ that is $4\times 4$ real antisymmetric.

In view of Remark \ref{FindingLog}
\[
Y = Y_{1} + Y_{2}
\]
with $[Y_{1}, Y_{2}]= 0$. Thus $\hat{Y} = \hat{Y_{1}} +
\hat{Y_{2}}$, where $\hat{Y_{l}} = Y_{l} \oplus 0_{1\times 1}, l= 1, 2$. Clearly
$\hat{Y_{1}}$ and $\hat{Y_{2}}$ also commute. Therefore 
\[
\Psi_{4,1}^{-1}(\hat{Y}) = \Psi_{4,1}^{-1}(\hat{Y_{1}}) +
\Psi_{4,1}^{-1}(\hat{Y_{2}})
\]
and as $\Psi_{4,1}$ is a Lie algebra isomorphism we find 
that the two summands on the right hand side of the last equation
also commute. Thus 
\[
{\mbox Exp} [\Psi_{4,1}^{-1}(\hat{Y}) ]
= {\mbox Exp} [\Psi_{4,1}^{-1}(\hat{Y_{1}})]
{\mbox Exp} [\Psi_{4,1}^{-1}(\hat{Y_{2}})].
\]

Now, the preimage of $V = {\mbox Exp}(\hat{Y})$ under $\Phi_{4,1}$ is $\pm 
{\mbox Exp} [\Psi_{4,1}^{-1}(\hat{Y})]$, which in turn     
is $\pm$ the product of 
the ${\mbox Exp} [\Psi_{4,1}^{-1}(\hat{Y_{l}})], \quad l=1, 2$.

Let us write $\Psi_{4,1}^{-1} (\hat{Y_{l}}) = 
\Theta_{{\mathbf H}} (Z_{l} + W_{l}j), \quad l=1,2.$
Then, evidently 
\[
{\mbox Exp} [\Psi_{4,1} (\hat{Y_{l}})]
= \Theta_{{\mathbf H}} [ {\mbox Exp} (Z_{} + W_{l}j], \quad l=1,2.
\]

Now both $Z_{l} + W_{l}j$ for $l=1,2$ satisfy a cubic polynomial
\[
(Z_{l} + W_{l}j)^{3} = -\kappa_{l}^{2} (Z_{l} + W_{l}j), \quad l=1,2.
\]
where $\kappa_{l}$ are real (as will be shown presently). Hence
\begin{equation}
\label{exp1}
{\mbox Exp} (Z_{l} + W_{l}j) = I_{2} + \frac{\sin (\kappa_{l})}{\kappa_{l}}
(Z_{l} + W_{l}j) + \frac{1-\cos (\kappa_{l})}{\kappa_{l}^{2}}
(Z_{l} + W_{l}j)^{2}
\end{equation}
and hence finding $\Phi_{4,1}^{-1}(V)$ (thus, $\Phi_{4,1}(X)$)
is complete.

We will next justify the claim that
$Z_{1} + W_{1}j$ is indeed annihilated by a real cubic
polynomial. First, inspecting Equation (\ref{Y1}) and Equation (\ref{ImPsi41}),
it is evident that we must also impose $a_{1} = 0, \beta_{2}
= \gamma_{2} = \alpha_{1}, b= -c = a_{2}, \beta_{1} = \gamma_{1} =
-\alpha_{2}$ in Equation (\ref{EquationZ}) and 
Equation (\ref{EquationW}) to obtain $Z_{1} + W_{1}j$.
This then yields
\[
Z_{1} = a_{2} \left (\begin{array}{cc}
i & 1\\
-1 & i
\end{array}
\right )
\]

and 
\[
W_{1} = \left ( \begin{array}{cc}
\alpha_{1} + \alpha_{2} & -\alpha_{2} + \alpha_{1}\\
-\alpha_{2} + \alpha_{1} & -\alpha_{1} - \alpha_{2}
\end{array}
\right ).
\]

Next, a direct calculation yields 
\[
(Z_{1} + W_{1}j)^{2} = (Z_{1}^{2} - W_{1}\bar{W_{1}}) + 
(Z_{1}W_{1} +W_{1}\bar{Z_{1}})j.
\]

A quick calculation then shows $Z_{1}W_{1} +W_{1}\bar{Z_{1}} = 0$ and 

\[
Z_{1}^{2} 
= 2a_{2}^{2} \left (\begin{array}{cr}
-1 & i\\
-i & -1 
\end{array}
\right )
\]

while
\[
W_{1}\bar{W_{1}}
= 2 (\alpha_{1}^{2} + \alpha_{2}^{2}) \left ( \begin{array}{cr}
1 & -i\\
i & 1
\end{array}
\right ).
\]
Thus $Z^{2} - W_{2}\bar{W_{2}}
= 2 (a_{2}^{2} + \alpha_{1}^{2} + \alpha_{2}^{2})
\left (\begin{array}{cr}
-1 & i\\
-i & -1 
\end{array}
\right )$.
Hence
\[
(Z_{1} + W_{1}j)^{3} = -4 (a_{2}^{2} + \alpha_{1}^{2} + \alpha_{2}^{2})
(Z_{1} + W_{1}j).
\]
In other words
\[
\kappa_{1} = 2(a_{2}^{2} + \alpha_{1}^{2} + \alpha_{2}^{2})^{1/2}.
\]
  
Similarly, $Z_{2}+ W_{2}j$ is expressible solely in terms of
$a_{2}, \alpha_{1}, \alpha_{2}$ (but, of course, the triple
$(a_{2}, \alpha_{1}, \alpha_{2})$ for $Z_{2} + W_{2}j$
is different from that for $Z_{1} + W_{1}j$).
Once again $(Z_{1} + W_{1}j)^{3} =  -\kappa_{2}^{2} (Z_{2} + W_{2}j)$,
with $\kappa_{2} = 2(a_{2}^{2} + \alpha_{1}^{2} + \alpha_{2}^{2})^{1/2}$.

This completes the inversion of $\Phi_{4,1}$ via the polar decomposition
which we present as an algorithm below.

\begin{algorithm}\label{Polarfor41}
{\rm 
\begin{itemize}
\item[1.] Let $X\in {\rm SO}^{+}(4,1)$. 
Compute using Remark \ref{FindingLog} both the polar decomposition $X=VP$ and the ``logarithms'' 
$Q\in so(4,1)$ of $P$ and the logarithm 
$\hat{Y} = (Y_{1} + Y_{2})\oplus 0_{1\times 1}$ in $so(4,1)$, of $V$, where 
$Y_{1}, Y_{2}$ are as in Equation (\ref{Y1}) and Equation (\ref{Y2}).
\item[2.] Find $\Lambda = \Psi_{4,1}^{-1}(Q)$ and $\lambda$ as given
by Equation (\ref{LambdaSymmetric}) and Equation (\ref{lambda}).
Then $\Phi_{4,1}^{-1} (P) = \pm \Theta_{{\mathbf H}} 
[\cosh (\lambda )I_{2} + \sinh (\lambda )\Lambda]$.
\item[3.] Next find $Z_{i} + W_{i}j\in M(2, {\mathbf H})$ and
$\kappa_{i}\in {\mathbf R}$ for $i=1,2$, from the entries 
of $Y_{1}, Y_{2}$. Then
\[
\Phi_{4,1}^{-1}(V) = \pm
\Theta_{{\mathbf H}}\{[ I_{2} + 
\frac{\sin (\kappa_{1}}{\kappa_{1}}(Z_{1} + W_{1}j)\\
+ \frac{1-\cos (\kappa_{1}}{\kappa_{1}^{2}} (Z_{1} + W_{1}j)^{2}]
[I_{2} + \frac{\sin (\kappa_{2}}{\kappa_{2}}(Z_{2} + W_{2}j)\\
+ \frac{1-\cos (\kappa_{2}}{\kappa_{2}^{2}} (Z_{2} + W_{2}j)^{2}]\}.
\]
\end{itemize}
}
\end{algorithm}

\begin{remark}\label{Polarin41}
{\rm As mentioned at the beginning of this section, the above considerations
can be used to compute the polar decomposition of a matrix $Y$ in
${\mbox Spin^{+}}(4,1)$, without computing that of the associated
$4\times 4$ complex matrix that is the $\Theta_{\mathbf{H}}$ image of it.
Indeed, all that one has to do 
is to compute $X = \Phi_{4,1}(Y)$ and apply 
the previous algorithm to $X$.}
\end{remark}

\section{Conclusions}\label{Section 7}

Explicit algorithms for inverting the double covering maps
$\Phi_{p, q}$, for $(p, q)\in \{(2,1)$, $(2,2)$, $(3,2)$, $(4,1)\}$
were provided. These methods extend for the general $(p, q)$
case, at the cost of more computation. 

A brief, and necessarily incomplete, comparison of the methods proposed here
and also the formula in \cite{shiro} follows.  Both our methods and
the method in \cite{shiro} will require considerable computation if
$n = p + q$ is large. Our methods require that a matrix form of
$\Phi_{p, q}$ be available first. This is, in any case, a necessity
if the principal aim of inversion is to relate matrix theoretic properties
of an element in the indefinite orthogonal group to those of its preimage(s)
in the spin group.On the other hand, this aim is, in general, impossible to
execute and achieve when viewing $\Phi_{p,q}$ only
as an abstract map.  Next, as $n$ grows the matrix entries of $\Phi_{p, q}(Y)$
will be quadratic entries in a large number of variables. On the other hand,
the formula in \cite{shiro} will require the computation 
of a prohibitive 
number of determinants. Next, among our methods, it is more direct to
use Groebner bases for inversion - if the matrix form of
$\Phi_{p, q}$ has been already calculated. 
The systems of equations and the attendant
the Groebner basis calculations become cumbersome if the polar decomposition
is used, instead of the Givens decompositions. This why in Section \ref{Section 3},
we did not use Groebner basis techniques. On the other hand, the
number of systems to be solved, when the Givens decomposition 
is employed, is larger than when the polar decomposition is deployed.
Note, however, the number of such systems to be solved symbolically,
in the Givens case,
is typically lower than the $\left (\begin{array}{c}
p+q\\
2
\end{array}
\right )$ Givens factors, since there is repetition of these different
factors (albeit with different $\theta$ or $\beta$) - see ii) of Remark \ref{However} in Section \ref{Section 2.4}. 
Finally, the Lie algebraic methods proposed require first that $\Psi_{p,q}$
be calcuated. This is no harder than finding the entries of $\Phi_{p,q}$, but it is
nevertheless a requisite.
The inversion of $\Psi_{p, q}$ is, of course, orders of magnitude simpler
than that of $\Phi_{p,q}$. However, to be able to use it effectively
for the inversion of $\Phi_{p, q}$,
one needs to be able to compute exponentials of matrices in the spin Lie
algebra easily. This factor is the basic tradeoff between Groebner
basis methods and the Lie algebraic method proposed here. For $n\leq 6$,
in most cases, there are explicit formulae for the exponential. The more
this can be extended to larger $n$, the Lie algebraic method becomes
more competitive.  On the other hand, for any $n$ the exponentiation is always
elementary when Givens factors are used as Remark \ref{agnostic} shows.
Finally, the combination of linearization and Givens factors provides an
alternative for the inversion of the abstract covering map and also what
is dubbed agnostic inversion of the concrete covering map, for any $n$.      
\section{Appendix I: Special Bases for Clifford Algebras}\label{Section 8}
In this appendix we show that every ${\mbox Cl}(p, q)$ possesses a basis
of one-vectors satisfying {\bf BP1} and {\bf BP 2} of Section \ref{Section 2.3}.
We note that the work, \cite{perti}, also provides special bases of
one-vectors for real Clifford algebras, but the properties of these
special bases are neither {\bf BP1} nor {\bf BP2}. 

We begin by recalling three iterative constructions for Clifford algebras,
\cite{pertii,portei} and show that these constructions inherit {\bf BP1}
and {\bf BP2}.

\begin{itemize}
\item {\bf IC1} If $\{V_{1}, \ldots , V_{p}, W_{1}, \ldots , W_{q}\}$
is a basis of one-vectors for ${\mbox Cl} (p, q)$ then
\[
\sigma_{z}\otimes V_{j}, \quad \left (\begin{array}{cc}
0 & I\\
I & 0
\end{array}
\right ),\quad \sigma_{z}\otimes W_{k}, \quad \left (\begin{array}{cc}
0 & I\\
-I & 0
\end{array}
\right )
\]
is a basis of one-vectors for ${\mbox Cl}(p+1, q+1)$. Here $I$ is the identity
element of ${\mbox Cl}(p, q)$ and $0$ is the zero element of ${\mbox Cl}(p, q)$.

Let $X\in \{V_{1}, \ldots , V_{p}, W_{1}, \ldots , W_{q}\}$. 
Then note that $ \left (\begin{array}{cc}
0 & I\\
I & 0
\end{array}
\right )^{*} [\sigma_{z}\otimes X]$ is a $2\times 2$ block matrix
with zeroes on its diagonal block. Similarly,
$ \left (\begin{array}{cc}
0 & I\\
-I & 0
\end{array}
\right )^{*} [\sigma_{z}\otimes X]$ also has zero diagonal blocks.
Similarly, the trace (respectively real part of trace) of 
$ \left (\begin{array}{cc}
0 & I\\
I & 0
\end{array}
\right )^{*}\left (\begin{array}{cc}
0 & I\\
-I & 0
\end{array}
\right )$ is also zero. Finally, if $X^{*}Y$ has zero trace 
(respectively zero real part of trace) then the same holds
for $(\sigma_{z}\otimes X)^{*}(\sigma_{z}\otimes Y)$.
So property {\bf BP2} is inherited by the iteration {\bf IC1}.
That property {\bf BP1} is also inherited by the iteration {\bf IC1}
is evident.   
\item {\bf IC2}
If $\{E_{1}, \ldots , E_{m}\}$ is a basis of
one-vectors for ${\mbox Cl}(m , 0)$ then the following set
is a basis of one-vectors for ${\rm Cl}(m + 8, 0)$
\[
\{ I\otimes V_{1}, \ldots , I\otimes V_{8},
	E_{1}\otimes L, \ldots , E_{}\otimes L\}
\]
where
\begin{itemize}
\item $I$ is the identity on ${\mbox Cl}(m, 0)$.
\item $\{V_{1}, \ldots, V_{8}\}$ is the basis of one-vectors for
${\mbox Cl}(8, 0)$ used in Theorem \ref{SpecialB} below. 
\item $L = \sigma_{x}\otimes \sigma_{x}\otimes  i\sigma_{y}\otimes
i\sigma_{y}$.
\end{itemize}

Note that $L$ is a real symmetric matrix. The $V_{i}$'s are also
real and either symmetric or antisymmetric.
Therefore, by item ii) of Remark \ref{sundry}
the reality of $L$ and the $V_{i}$
ensures that {\bf BP2} is inherited by {\bf IC2}.
Since $L$ is real symmetric  and $V_{i}^{T}
= \pm V_{i}$, we also see that {\bf BP1} is also inherited
by {\bf IC2}.
\item {\bf IC3} If $\{F_{1}, \ldots , F_{m}\}$ is a basis of one-vectors
for ${\mbox Cl}(0, m)$ then the following is a basis of 
one-vectors for ${\mbox Cl}(0, m+ 8)$
\[
\{ I\otimes V_{1}, \ldots , I\otimes V_{8},
F_{1}\otimes K, \ldots , F_{m}\otimes K\}
\]
where
\begin{itemize}
\item $I$ is the identity on ${\mbox Cl}(0, m)$.
\item $\{V_{1}, \ldots, V_{8}\}$ is the basis of one-vectors for
${\mbox Cl}(0, 8)$ used in Theorem \ref{SpecialB} below. 
\item $K = i\sigma_{y}\otimes \i\sigma_{y}\otimes  \sigma_{z}\otimes
\sigma_{z}$.
\end{itemize}
As in the previous case $K$ is real symmetric, while
each $V_{i}$ is real and either symmetric or antisymmetric. Therefore,
both {\bf BP1} and {\bf BP2} are inherited by {\bf IC3}.
 
\end{itemize}
 
We are now in a position to prove the main result of this appendix.

\begin{theorem}\label{SpecialB}
Every real Clifford algebra has a basis of one-vectors with the properties
{\bf BP1} and {\bf BP2}.
\end{theorem}
 
\noindent {\bf Proof:}
As observed above both {\bf BP1} and {\bf BP2} are inherited by each
of {\bf IC1}, {\bf IC2} and {\bf IC3}. 
Since every ${\mbox Cl}(r, s)$ can be obtained by
repeatedly applying {\bf IC1} to either some ${\mbox Cl}(n, 0)$ or
${\mbox Cl}(0, n)$, and every ${\mbox Cl}(n, 0)$ (respectively 
${\mbox Cl}(0, n)$)
is obtained by applying {\bf IC2} (respectively {\bf IC3})
to ${\mbox Cl}(m, 0), m=0,\ldots, 8$ (respectively ${\mbox Cl}(0, m), m=0,\ldots, 8$)
it suffices to verify the theorem for ${\mbox Cl}(m, 0)$ and ${\mbox Cl}(0, m)$
for $m= 0, \ldots , 8$. 

Let us begin with ${\mbox Cl}(m, 0)$.
The following is the list of
bases of one-vectors that will be used for this purpose

\begin{itemize}
\item $B_{0,0} = \Phi$.
\item $B_{1,0} = \{\sigma_{x}\}$.
\item $B_{2,0} = \{\sigma_{z}, \sigma_{x}\}$. 
\item $B_{3.0} = \{\sigma_{z}, \sigma_{x}, i\sigma_{y}\}$.
\item $B_{4,0} = \{\left (\begin{array}{cc}
0 & i\\
-i & 0
\end{array}
\right ), \left (\begin{array}{cc}
0 & j\\
-j & 0
\end{array}
\right ), \left (\begin{array}{cc}
0 & k\\
-k & 0
\end{array}
\right ), \sigma_{z}\}$.
\item 
$B_{5,0} =\{\sigma_{x}\otimes\sigma_{z}\otimes (i), \sigma_{x}\otimes
\sigma_{z}\otimes (j), \sigma_{x}\otimes\sigma_{z}\otimes (k), 
\sigma_{z}\otimes\sigma_{z},\sigma_{x}\otimes\sigma_{z}\}$.
\item $B_{6,0} =$ $\{ I_{2}\otimes \sigma_{z}$, $I_{2}\otimes \sigma_{x}$,
$i I_{2}\otimes (i\sigma_{y})$, $j I_{2}\otimes (i\sigma_{y})$,
$k\sigma_{x}\otimes (i\sigma_{y})$, $k\sigma_{z}\otimes (i\sigma_{y})\}$.
\item $B_{7,0} =\{I_{4}\otimes \sigma_{z}$, $I_{4}\otimes \sigma_{x}$,
$-i\sigma_{z}\otimes I_{2}\otimes (i\sigma_{y})$,
$i\sigma_{y}\otimes I_{2}\otimes (i\sigma_{y})$,
$-i\sigma_{x}\otimes \sigma_{x}\otimes (i\sigma_{y})$,
$-i\sigma_{x}\otimes \sigma_{z}\otimes (i\sigma_{y})$,
$\sigma_{x}\otimes -i\sigma_{y}\otimes (i\sigma_{y})\}$.
\item $B_{8,0}=\{ I_{8}\otimes \sigma_{z}$, $I_{8}\otimes \sigma_{x}$,
$-\sigma_{x}\otimes i\sigma_{y}\otimes I_{2}\otimes (i\sigma_{y})$,\\
$-i\sigma_{y}\otimes I_{2}\otimes I_{2}\otimes (i\sigma_{y})$,
$-\sigma_{z}\otimes i\sigma_{y}\otimes \sigma_{z}\otimes (i\sigma_{y})$,
$-\sigma_{z}\otimes \i\sigma_{y}\otimes \sigma_{x}\otimes (i\sigma_{y})$,
$\sigma_{z}\otimes I_{2} \otimes i\sigma_{y}\otimes (i\sigma_{y})$,
$-\sigma_{x}\otimes\sigma_{z}\otimes i\sigma_{y}\otimes (i\sigma_{y})\}$
$=\{V_{1}, \ldots , V_{8}\}$.
\end{itemize}
By construction {\bf BP1} and {\bf BP2} hold for these eight bases.
Next we verify the assertion for ${\mbox Cl}(0, m)$. We work the following
sets of one-vectors for $m\leq 8$

\begin{itemize}
\item $B_{0,1} = \{i\}$.
\item $B_{0,2} = \{i, j\}$.
\item $B_{0, 3} = \{ \left (\begin{array}{cc}
	i & 0\\
0 & i
\end{array}
\right ), \left (\begin{array}{cc}
j & 0\\
0 & j 
\end{array}
\right ), \left (\begin{array}{cc}
k & 0\\
0 & k 
\end{array}
\right ) \}$.

\item $B_{0, 4} = \{ \left (\begin{array}{cc}
i & 0\\
0 & i
\end{array}
\right ), \left (\begin{array}{cc}
j & 0\\
0 & j 
\end{array}
\right ), \left (\begin{array}{cc}
k & 0\\
0 & k 
\end{array}
\right ), \left (\begin{array}{cc}
0 & k\\
k & 0 
\end{array}
\right ) \}$. 

\item $B_{0, 5}
= \{ i\sigma_{z}\otimes I_{2}, i\sigma_{y}\otimes I_{2},
i\sigma_{x}\otimes \sigma_{x}, i\sigma_{x}\otimes \sigma_{z},
\sigma_{x}\otimes (i\sigma_{y})\}$.

\item $B_{0,6}
= \{\sigma_{z}\otimes (i\sigma_{y})\otimes I_{2}, 
i\sigma_{y}\otimes I_{4}, \sigma_{x}\otimes (i\sigma_{y})\otimes\sigma_{x},
 \sigma_{x}\otimes (i\sigma_{y})\otimes\sigma_{z}, \ \\ 
\sigma_{x}\otimes I_{2}\otimes I_{2} \otimes (i\sigma_{y}),
\sigma_{z}\otimes\sigma_{x}\otimes (i\sigma_{y})\}
= \{ Z_{1}, \ldots , Z_{6}\}$.

\item $B_{0,7}= \{Z_1\oplus Z_1,\ldots, Z_6\oplus Z_6, \sigma_{z}\otimes 
\sigma_{z}\otimes (i\sigma_{y})\oplus\sigma_{z}\otimes \sigma_{z}\otimes 
(i\sigma_{y})\}$, where the $Z_j$'s are defined as in the previous item.

\item $B_{0, 8} = \{ I_{4}\otimes \sigma_{z}\otimes (i\sigma_{y}),
I_{4}\otimes (i\sigma_{y})\otimes I_{2}, I_{2}\otimes \sigma_{z}\otimes
\sigma_{z}\otimes (i\sigma_{y}), I_{2}\otimes\sigma_{z}\otimes\sigma_{x}
\otimes (i\sigma_{y}), I_{2}\otimes (i\sigma_{y})\otimes\sigma_{x}\otimes
I_{2}, I_{2}\otimes (i\sigma_{y})\otimes \sigma_{z}\otimes\sigma_{x},
\sigma_{x}\otimes (i\sigma_{y})\otimes \sigma_{z}\otimes \sigma_{z},
\sigma_{z}\otimes (i\sigma_{y})\otimes \sigma_{z}\otimes \sigma_{z}\}=
\{V_{1}, \ldots , V_{8}\}$.

\end{itemize}

Again, by construction {\bf BP1} and {\bf BP2} hold for these bases also.
This concludes the proof.

\end{document}